\documentclass[aps,twocolumn,notitlepage,superscriptaddress]{revtex4-1} 
\usepackage{times,color,amsfonts,amssymb,amsbsy,amsthm,amsmath,graphics,graphicx,hyperref,bm,bbm,makecell,mathtools,dsfont,upgreek,hyperref,fancyhdr,multirow,braket}
\usepackage{blindtext}
\usepackage[capitalise]{cleveref}
\usepackage{tcolorbox} 
\hypersetup{colorlinks,linkcolor={black},citecolor={black},urlcolor={black}}
\definecolor{psred}{rgb}{0.8,0.2,0.2}
\newcommand{\MG}[1]{\textcolor{black}{#1}}
\newcommand{\SP}[1]{\textcolor{black}{#1}}
\newcommand{\GS}[1]{\textcolor{black}{#1}}
\begin{document}
\title{An Overview of CV-MDI-QKD}
%\title{An Overview on Continuous-Variable \\Measurement-Device-Independent Quantum Key Distribution}
\author{Alasdair I. Fletcher}
\affiliation{Department of Computer Science, University of York, York YO10 5GH, United Kingdom}
\affiliation{nodeQ, 71-75 Shelton Street, London WC2H 9JQ, United Kingdom}

\author{Cillian Harney}
\affiliation{Department of Computer Science, University of York, York YO10 5GH, United Kingdom}
\affiliation{nodeQ, 71-75 Shelton Street, London WC2H 9JQ, United Kingdom}

\author{Masoud Ghalaii} 
\affiliation{Department of Computing and Mathematics, Manchester Metropolitan University, Manchester M1 5GD, United Kingdom}

\author{Panagiotis Papanastasiou}
\affiliation{Department of Computer Science, University of York, York YO10 5GH, United Kingdom}
\affiliation{nodeQ, 71-75 Shelton Street, London WC2H 9JQ, United Kingdom}
\affiliation{Department of Physics, University of York, York YO10 5DD, United Kingdom}

\author{Alexandros Mountogiannakis}
\affiliation{nodeQ, 71-75 Shelton Street, London WC2H 9JQ, United Kingdom}

\author{Gaetana Spedalieri}
\affiliation{Department of Computer Science, University of York, York YO10 5GH, United Kingdom}

\author{Adnan A. E. Hajomer}
\affiliation{Department of Physics, Technical University of Denmark, 
2800 Kongens Lyngby, Denmark}
\author{Tobias Gehring}
\affiliation{Department of Physics, Technical University of Denmark, 2800 Kongens Lyngby, Denmark}

\author{Stefano Pirandola}
\thanks{Corresponding author}
\email{stefano.pirandola@york.ac.uk}
\affiliation{Department of Computer Science, University of York, York YO10 5GH, United Kingdom}

\begin{abstract}
As quantum key distribution (QKD) emerges as a robust defense against quantum computer threats, significant advancements have been realized by researchers. A pivotal focus has been the development of protocols that not only simplify hardware implementation, such as the use of continuous-variable (CV) systems, but also eliminate the necessity for trusted nodes, as seen with the measurement-device-independent (MDI) approach. This paper delves into the integration of these methodologies in the CV-MDI-QKD protocol, offering an in-depth exploration of its evolution, primary characteristics, and the latest advancements in both theory and experiment.
\end{abstract}

\flushbottom
\maketitle
\thispagestyle{empty}

%%%%%%%%%%%%%%%%%%%%%%%%%% SHORTER INTRO %%%%%%%%%%%%%%%%%%%%%

\section{Introduction}

Four decades have passed since quantum-secure communications were introduced~\cite{BB84,BB84:2004}. 
% Besides quantum cryptography~\cite{Pirandola:AQCrypt2020}, the wider tree of quantum communications has branched out to several distinct areas of research, including quantum repeaters~\cite{Briegel:QRs1998,Dur:QRs1999,Duan:QRs2001,Azuma:QRs2017}, quantum networks~\cite{Elliott:QNet2002,Xu:QNet2009,Peev:QNet2009,Sasaki:QNet2011,Razavi:Book2018,Epping:QNet2016,Pirandola:QNet2019a,Pirandola:QNet2019b} and quantum memories~\cite{Clausen:QMs2011,Saglamyurek:QMs2011,Sangouard:QMs2011,Sinclair:QMs2014}. 
Besides quantum key distribution (QKD)~\cite{Pirandola:AQCrypt2020}, we have also witnessed the birth of other quantum-secure technologies, including quantum random number generators~\cite{Colbeck:QRNGs2012,Gallego:QRNGs2013,Mitchell:QRNGs2015,Lunghi:QRNGs2015,Brandao:QRNGs2016,Ma:QRNGs2016,Acin:QRNGs2016,Herrero-Collantes:QRNGs2017}, quantum conferencing or secret-sharing~\cite{Hillery:QSS1999,Cleve:QSS1999,Ribeiro:QSS2010,Keet:QSS2010,Ottaviani:QConf2019,Grasselli:QSS2019,Pirandola:QConf2020}, quantum digital signatures~\cite{Gottesman:QDS2001,Clarke:QDS2012,Collins:QDS2014,Dunjko:QDS2014,Wallden:QDS2015,Croal:QDS2016,Collins:QDS2016,Roberts:QDS2017}, quantum bit commitment~\cite{Mayers:BitCom1997,LO:BitCom1998,Buhrman::BitCom2008,Chailloux:BitCom2011,Kent:BitCom2012}, and secure quantum computing~\cite{Broadbent:SQC2009,Barz:SQC2012,Barz:SQC2013,Fitzsimons:SQC2017,Kashefi:SQC2017}, to name a few.
%, as well as quantum error correction techniques for entanglement distillation~\cite{Jiang:QDist2009,Munro:QDist2010,Campbell:QDist2013,Dias:QDist2018,Seshadreesan:QDist2019}. 
These outstanding endeavours are contributing to the future quantum internet~\cite{Kimble:QInt2008,Azuma:QNet2016,Pirandola:QNet2016,Barz:QNet2018,Azuma:QInt2019}. 

%and have made closing several loopholes in the Bell inequality experiments possible~\cite{Giustina:LoopBell2015,Shalm:LoopBell2015,Hensen:LoopBell2015}.

% Among these technologies, quantum key distribution (QKD) stands as the most mature~\cite{Pirandola:AQCrypt2020}, and has been proven a dynamic research area undergoing rapid theoretical and experimental progress. 

There exist two major classes of QKD, i.e., discrete-variable (DV) and continuous-variable (CV) protocols. The former class exploits quantum resources with finite degrees of freedom, e.g.,~polarization qubits, and contains the BB84~\cite{BB84}, E91~\cite{Ekert:1991}, B92~\cite{Bennett:1992} and six-state~\cite{Bruss:SixSt1998} protocols.
%and measurement-device-independent (MDI) protocols~\cite{Braunstein:MDI2012,Lo:MDI2012}. 
In contrast, CV-QKD encodes information into quadratures of light~\cite{Weedbrook:RMP2012} representing the so-called `continuous variables'. CV-QKD forms a distinct class of communication strategies including those which are based on squeezed states~\cite{Ralph:CVQKD-Sq1999,Hillery:CVQKD-Sq2000,Reid:CVQKD-Sq2000},
coherent states~\cite{Grosshans:PRL2002,Grosshans:Nat2003}, and noisy/thermal states either in optical~\cite{Filip:CVQKD-Th2008,Usenko:CVQKD-Th2010,Weedbrook:CVQKD-Th2010,Weedbrook:CVQKD-Th2012} or non-optical~\cite{Weedbrook:CVQKD-Th2014,Ottaviani:CVQKD-THz2020} regimes. 
Other techniques continue to be explored involving no-switching~\cite{Weedbrook:CVQKD2004}, discrete modulation~\cite{Leverrier:DMCVQKD2009}, and two-way~\cite{Pirandola:CV2Way2008} CV-QKD protocols. 
It is also worth to mention hybrid protocols where both DV and CV components are exploited~\cite{Djordjevic:hybridQKD20,Djordjevic:hybridQKD22,Primaatmaja:hybridQKD,Mycroft:hybridQKD}.

All QKD protocols mentioned above are endowed with their own unique properties, and their success is based upon several salient features. This includes (\textit{i}) their underlying trade-off between rate and transmission distance, (\textit{ii}) their strength of security, and (\textit{iii}) their feasibility of practical deployment. The theoretical and experimental evolution of QKD over the years has aimed to accelerate progress in each of these directions, with several critical milestones being reached along the way. 

To address (\textit{i}), one such milestone was the derivation of the Pirandola-Laurenza-Ottaviani-Banchi (PLOB) bound~\cite{Pirandola:PLOB2017} which establishes the exact fundamental trade-off between communication rate and distance. This limit puts an upper bound on all point-to-point implementations of QKD. Namely, for a lossy bosonic channel, which could underlie models of optical-fibre and free-space channels, two users cannot distribute more than the secret key capacity of the link $-\log_2(1-\tau)$, where $\tau$ quantifies channel's loss. The PLOB bound provides a guide for all point-to-point QKD protocols, outlining an optimal performance level that can guide future protocol/experimental designs. 

The PLOB bound also emphasises the critical inability of repeaterless, high-rate quantum communication over long-distance channels~\cite{PirandolaPatron2009, Pirandola:PLOB2017}. However, realistic communication networks would consist of many interconnected nodes, and parties would not rely merely on direct links. Hence, this innate weakness can be compensated by additional infrastructure in the form of repeaters and quantum networks. With the addition of intermediate stations, distance limitations can be circumvented and high-rates restored. Indeed, the PLOB bound has more recently been extended to derive end-to-end capacities of quantum-safe networks \cite{Pirandola:QNet2019a,Pirandola:QNet2019b}. These results derive critical bounds on the performance of repeater-assisted QKD protocols, motivating advances within this domain. 

Nevertheless, the development of repeater-assisted protocols invites further questions. In Refs.~\cite{Pirandola:QNet2019a,Pirandola:QNet2019b}, the end-to-end capacities are shown to be achievable by assuming either trusted QKD repeaters or entanglement-based repeaters. In a practical QKD network, it would be ideal to use repeaters simultaneously \emph{untrusted}, so they could still be usable even if compromised by an attacker, and \emph{not based on entanglement distribution}, so they can be more easily deployed in realistic technological environments. 
\MG{ To address (\textit{ii}) and (\textit{iii}) simultaneously,} the first step done in this direction was the introduction of measurement device independent (MDI) QKD~\cite{Braunstein:MDI2012,Lo:MDI2012,Liu:MDI2012,Curty:MDI2014} showing that QKD is still possible in a single-hop chain where Alice and Bob connects to an untrusted, and entanglement-free, repeater. More recently, the MDI idea was extended to design the twin-field QKD protocol~\cite{Lucamarini:TF2018,Minder:TF2019}, which can also exceed the PLOB bound.

Soon after the introduction of the MDI protocol, a CV version was designed and experimentally realized~\cite{Pirandola:CVMDI2013, Pirandola:CVMDI2015}. CV-MDI-QKD aims to achieve the same features as DV-MDI-QKD while ensuring much higher key rates at short distances. Many works followed proposing \GS{other studies,} variants and improvements. 
\GS{Following Ref.~\cite{Pirandola:CVMDI2013}, CV-MDI-QKD was also investigated in Refs.~\cite{PhysRevA.89.042335,PhysRevA.89.052301} with some inaccuracies.} 
Several works implemented techniques based on photon subtraction~\cite{Zhao:CVMDI-subt2018,Ma:CVMDI-subt2018}, noiseless linear amplifiers~\cite{Zhang:CVMDI-NLA2015}, phase-sensitive amplifiers~\cite{Wang:CVMDI-NLA2019}\MG{, and in free-space settings~\cite{GhalaiiPRA2023}.} Attempts have been made on the encoding side by replacing coherent states with squeezed or thermal states~\cite{Zhang:CVMDI-SqSt2014,Bai:CVMDI-ThSt2019,Bai:CVMDI-UD2020}, and replacing Gaussian modulation with discrete modulation or dual-phase modulation~\cite{Ma:CVMDI-DM2019,Liao:CVMDI-DPhM2018}. %Focus was also put on the phase reference beam and alignment~\cite{Ma:CVMDI-PhRef2019,Yin:CVMDI-SelfAli2019}, as well as on relay concatenation~\cite{Guo:CVMDI2017}. 

Early works that assumed the asymptotic rates with many uses of the communication links have been extended to the composable and finite-size framework~\cite{Papanastasiou:FSCVMDI2017,Mountogiannakis:CompFS2022} which significantly strengthen the security analysis of such protocols for practical deployment. \GS{Post-selection} strategies for CV-MDI-QKD have also been recently investigated~\cite{Wilkinson:CVMDI2020} which can help to extend transmission distances. Furthermore, investigations have been carried out concerning how CV-MDI-QKD protocols can be integrated into more complex networking domains with more than two users \cite{Ottaviani:QConf2019, Fletcher:3nodeCVMDI2022} and more intermediate, untrusted nodes \cite{Ghalaii:GQNets2022}.

% This review is structured as follows. In Sec.~\ref{sec:Robust&Rate}, we discuss the robustness and secret key rate of CV-MDI-QKD under Gaussian attacks. We also talk through the use of post-selection to enhance these protocols. Sec.~\ref{sec:Comp&FS}, reviews finite-size and composibility, as well as error correction. In Sec.~\ref{sec:NetComplex}, we weigh up quantum network complexity, including CV-MDI-QKD star and three-node networks, as well as general Gaussian quantum networks with untrusted relays.  In Sec.~\ref{sec:experiments} we discuss experimental implementations of CV-MDI-QKD. We summarise our manuscript in Sec.~\ref{sec:conclusions}. There are several appendices which include details of some of the calculations.

\begin{figure*}
\vspace{+.1cm}
\includegraphics[width=11cm]{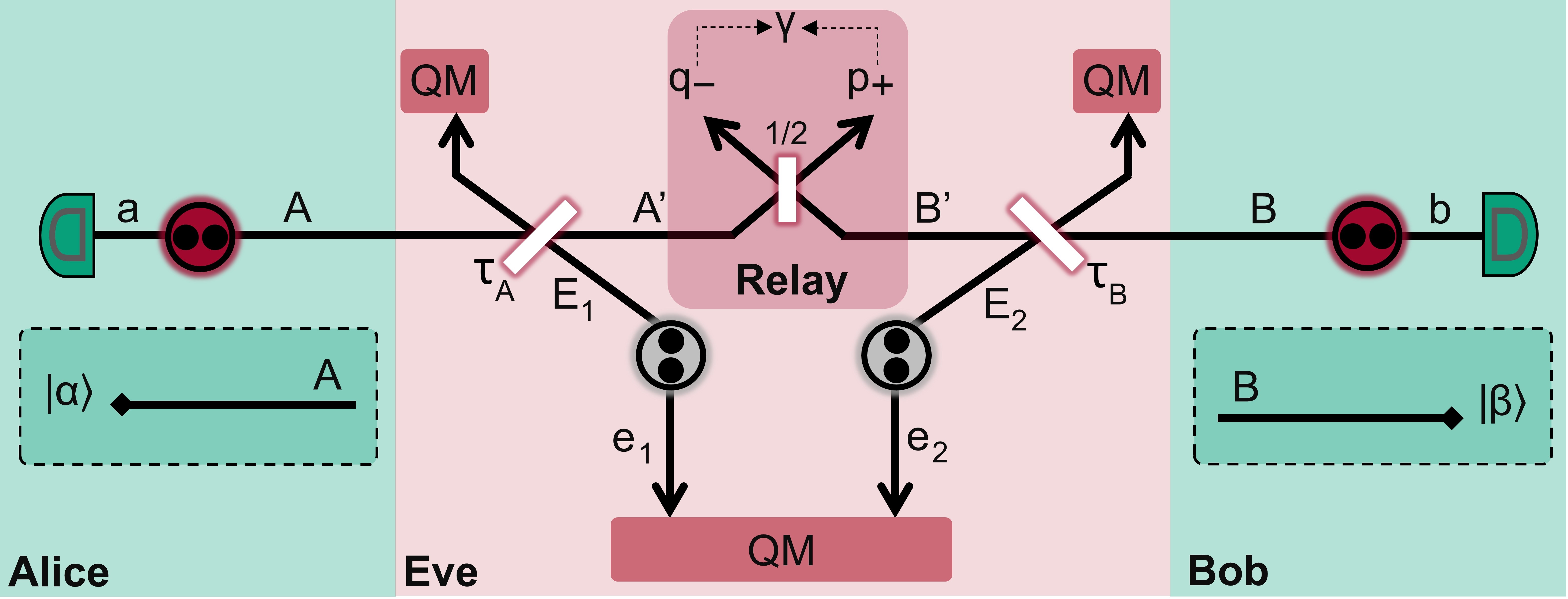}
\caption{
{\bf Schematic of CV-MDI-QKD.} In the entanglement-based description, Alice and Bob transmit one mode ($A$ and $B$) of their TMSV states, denoted by red circles. In the equivalent and practical prepare-and-measure description (dashed boxes), Alice and Bob directly send Gaussian-modulated coherent states to the relay. 
Eve has a pair of TMSV states, denoted by grey circles. Her modes $E_1$ and $E_2$ are made to interact with Alice and Bob’s signal modes via beam splitters of transmissivities $\tau_A$ and $\tau_B$, respectively (each of these interactions corresponds to an `entangling cloner'). Eve's output modes after the beamsplitters and her retained modes ($e_1$ and $e_2$) are stored in a quantum memory \GS{(QM)} for collective measurement at the end of the protocol (this is known as a `collective attack'). Alice's and Bob's output modes $A'$ and $B'$ are mixed in a $1/2$ beam splitter at the relay, whose outputs are subsequently measured using two conjugate homodyne detections with outcomes $q_-$ and $p_+$, respectively. These real outcomes form a single complex value $\gamma$ that is publicly broadcast.} %Reproduced from~\cite{Pirandola:CVMDI2015}.} 
\label{fig:CVQKD}
\end{figure*}

This review is structured as follows. In Sec.~\ref{features} we provide a schematic overview of the protocol. In Sec.~\ref{sec:Robust&Rate}, we discuss the secret key rate of CV-MDI-QKD based on Gaussian modulation and its post-selection variant. In Sec.~\ref{sec:Comp&FS}, we review finite-size and composability.  \GS{In Sec.~\ref{sec:NetComplex}, we weigh up quantum network complexity, including CV-MDI-QKD star and three-node networks}. In Sec.~\ref{sec:experiments} we discuss the main experimental implementations of CV-MDI-QKD. We summarize our manuscript in Sec.~\ref{sec:conclusions}.

\section{Features of the protocol}\label{features}

The basic protocol for CV-MDI-QKD~\cite{Pirandola:CVMDI2013, Pirandola:CVMDI2015} is sketched in Fig.~\ref{fig:CVQKD}. The steps are as follows:

{\bf a.}	
In the so-called entanglement-based (EB) representation, Alice and Bob each heterodyne one mode of a two-mode squeezed vacuum (TMSV) state while transmitting the other mode. This is equivalent to a prepare-and-measure (P\&M) scheme, where they prepare a coherent state and modulate its amplitude via a bi-variate Gaussian distribution. It is assumed that the labs of Alice and Bob are secure so the loss and noise within them are trusted. 

{\bf b.} 
Alice and Bob send their Gaussian-modulated coherent states $|\alpha\rangle$ and $|\beta\rangle$ to the relay through quantum channels that could be, e.g., optical fibres or free-space links. These channels \GS{can be} simulated via beam splitters with transmissivities $\tau_A$ and $\tau_B$. 
The eavesdropper (Eve) can attack the links by overlapping one mode of her TMSV states at the beam splitters. \GS{This `entangling-cloner attack' represents the most relevant Gaussian attack}. See the next section for more details.  

{\bf c.} 
At the relay, Charlie uses a CV Bell measurement to perform a joint detection. This step includes overlapping signals received from Alice and Bob’s station on a balanced beam splitter and homodyning both outlet beams. Precisely, Charlie measures orthogonal quadratures $q$ and $p$ obtaining $q_-$ and $p_+$, respectively. The resultant complex value $\gamma=q_-+ip_+$ is then broadcast to Alice and Bob. \GS{The outcome $\gamma$ establishes a-posteriori correlations between the variables of the parties. Consequently, knowing $\gamma$ allows each party to infer the variable of the other party through post-processing.}

{\bf d.} 
Alice and Bob post-process their measured/modulation data along with the corresponding relay outcomes, \GS{with the effect of correlating their data. Next, the} post-processing includes channel parameter estimation (PE), error correction (EC), and privacy amplification (PA). These operations will provide the final secret shared key.

\section{Security analysis}
\label{sec:Robust&Rate}
The two end users, Alice and Bob, each have access to independent, identical zero-mean bi-variate Gaussian distributions with the same variance $\mu -1$ (assumed to be large). From these distributions, they draw two amplitude values $\alpha$ and $\beta$ therefore preparing Gaussian-modulated coherent states $\ket{\alpha}$ and $\ket{\beta}$ respectively (see prepare-and-measure description in Fig.~\ref{fig:CVQKD}). These two states are sent to a central detection station, which is in principle controlled by Eve. Operated correctly, the station performs a Bell detection in which the states are combined on a balanced beamsplitter and undergo conjugate homodyne detections. The results of these measurements are combined into a single complex variable $\gamma$ which is publicly broadcast. It is clear that initially the mutual information $I(\alpha:\beta)=0$ since the coherent states are prepared from independent Gaussian distributions. However after the announcement of $\gamma$, Alice and Bob are able to decode each other's variable and thus $I(\alpha:\beta|\gamma)>0$. Despite having access to $\gamma$, Eve is unable to infer the parties' variables, i.e., $I(E:\alpha)=I(E:\beta)=0$. Eve is therefore forced to attack the channels or alter the operation of the relay, both of which may be detected by Alice and Bob in post-processing.

Let us now consider the same protocol in the equivalent entanglement-based representation \cite{Pirandola:AQCrypt2020}. A detailed discussion of the formulation can be found in the supplementary information of \cite{Pirandola:CVMDI2015}. Alice begins with a two-mode squeezed vacuum (TMSV) $\Phi_{aA}$, a zero mean Gaussian state with covariance matrix given by \cite{Weedbrook:RMP2012}:
\begin{equation}
\boldsymbol{V}=\begin{pmatrix}
\mu \boldsymbol{I} & \sqrt{\mu^2-1} \boldsymbol{Z} \\
\sqrt{\mu^2-1} \boldsymbol{Z} & \mu \boldsymbol{I}
\end{pmatrix}   
\label{eq:TMSV}
\end{equation}
where $\boldsymbol{Z}=\text{diag}(1,-1)$.  Mode $a$ undergoes heterodyne detection, projecting the travelling mode $A$ into the coherent state $\ket{\alpha}$, the modulation of which is in one-to-one correspondence with the measurement outcome  $\widetilde{\alpha}$. In the limit of large modulation $\mu \rightarrow \infty$, we have $\alpha=\widetilde{\alpha}^{*}$. The one-to-one correspondence between $\alpha$ and $\widetilde{\alpha}$ renders the two variables informationally equivalent. Symmetric description holds for Bob. 
Alice's and Bob's modes $A$ and $B$ are sent to the central relay for Bell detection and, by the commutation of local measurements, Alice's and Bob's heterodyne detections of their retained modes $a$ and $b$ may be delayed until the end of the protocol. The protocol can thus be seen as an entanglement swapping protocol, generating an entangled Gaussian state between Alice's and Bob's retained modes by the Bell detection of the transmitted modes. 
%We may therefore analyze the performance of the protocol by considering the mutual information between Alice and Bob's heterodyne detection variables $\widetilde{\alpha}_A$ and $\widetilde{\alpha}_B$. 
The entanglement-based representation is a useful mathematical tool for studying the security of the protocol but also provides a potential version of it that can prevent side-channel attacks on state preparation \cite{Braunstein:MDI2012}. 

\subsection{Gaussian attacks and asymptotic key rate}
Eve's most general attack strategy against an MDI protocol can always be reduced to the case in which the detector is operated correctly and Eve attacks the incoming links \cite{Pirandola:CVMDI2015}. Failing to undertake detections or measuring other observables can always be detected by the parties in post-processing \cite{Braunstein:MDI2012}. Moreover, the most powerful collective attack that Eve may undertake to eavesdrop on a Gaussian CV protocol is a Gaussian attack \cite{GarciaPatron:Optim2006}. Alice and Bob therefore assume the worst-case scenario where all the pure loss and thermal noise can be attributed to Eve performing such an optimal Gaussian attack on the incoming links. Such an attack may be modeled as mixing Alice and Bob's incoming states with Eve's ancillary states $E_1$ and $E_2$ at beamsplitters of transmissivity $\tau_1$ and $\tau_2$ respectively. The reduced state $\hat{\rho}_{E_1E_2}$ is a correlated thermal state with zero mean and covariance matrix given by:
\begin{equation}
\boldsymbol{V}_{E_1E_2}=\begin{pmatrix}
\omega_A \boldsymbol{I} & \boldsymbol{G}\\
\boldsymbol{G} & \omega_B \boldsymbol{I}
\end{pmatrix}
\end{equation}
where $\boldsymbol{G}=\text{diag}(g,g')$ such that $g$ and $g'$ are correlation parameters and $\omega_A$ and $\omega_B$ are the thermal variances introduced into each incoming link. For simplicity, one can study the case of \SP{$g'=g=0$}, which corresponds to the dilation into a pair of independent entangling cloners as depicted in Fig.~\ref{fig:CVQKD}. \SP{(We assume this case in the numerical plots of this paper).}

We may compute the mutual information between Alice and Bob's variables $I_{AB}:=I(\alpha:\beta)$ while Eve's total stolen information $I_E$ is upper bounded by using the Holevo information.
In the asymptotic limit of an infinite number of protocol rounds, the rate takes the form $R=I_{AB}-I_E$, so that Alice and Bob share on average $R$ secret bits per use of the relay. Note that the rate above is a simplified asymptotic formula that also assumes unit reconciliation efficiency (i.e., perfect error correction during data post-processing).
%In the finite case, we need to consider the reconciliation efficiency $\xi\leq1$:
%\begin{equation}\label{assymptotic_rate}
%R=\xi I_{AB}-I_E
%\end{equation}

\begin{figure}[h]
    \centering
    \includegraphics[width=\linewidth]{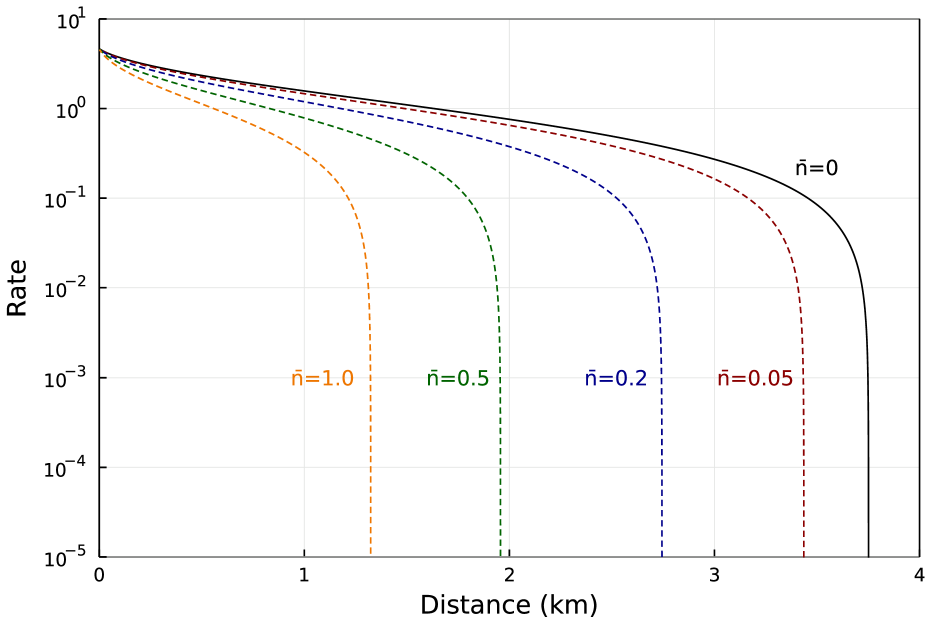}
    \caption{Asymptotic key rate (bits/use) of the symmetric CV-MDI-QKD protocol versus fibre distance ($0.2$dB/km) for various values of thermal noise $\bar{n}$ affecting each link. The solid black line indicates the rate under pure loss. All rates are optimized with large modulation. Unit reconciliation and perfect detection efficiency are assumed. \SP{ Figure adapted from Ref.~\cite{Fletcher:thesis2023}}.}
    \label{fig:2 user rate basic CV MDI}
\end{figure}

In the symmetric configuration, Alice and Bob are equidistant from the relay. In this case, the rate-distance scaling is shown in \cref{fig:2 user rate basic CV MDI}. The distance is calculated by assuming the parties connected to the relay by fibre optic links with a loss of $0.2$ dB/km. That is the transmissivity of the beamsplitters used in Eve's attack may be given by \MG{$\tau_i=10^{-0.02d_i}$ where $d_i$ is the distance in kilometers of the links to Eve.}

In the symmetric configuration, the protocol can achieve high rates at short distances ($\sim$1 bit per use) and is robust against thermal noise. However, the maximum distance where a positive rate can be maintained is limited. Under pure loss, a positive rate can only be kept for $\leq 4$km. 

\begin{figure}[h]
    \centering
    \includegraphics[width=\linewidth]{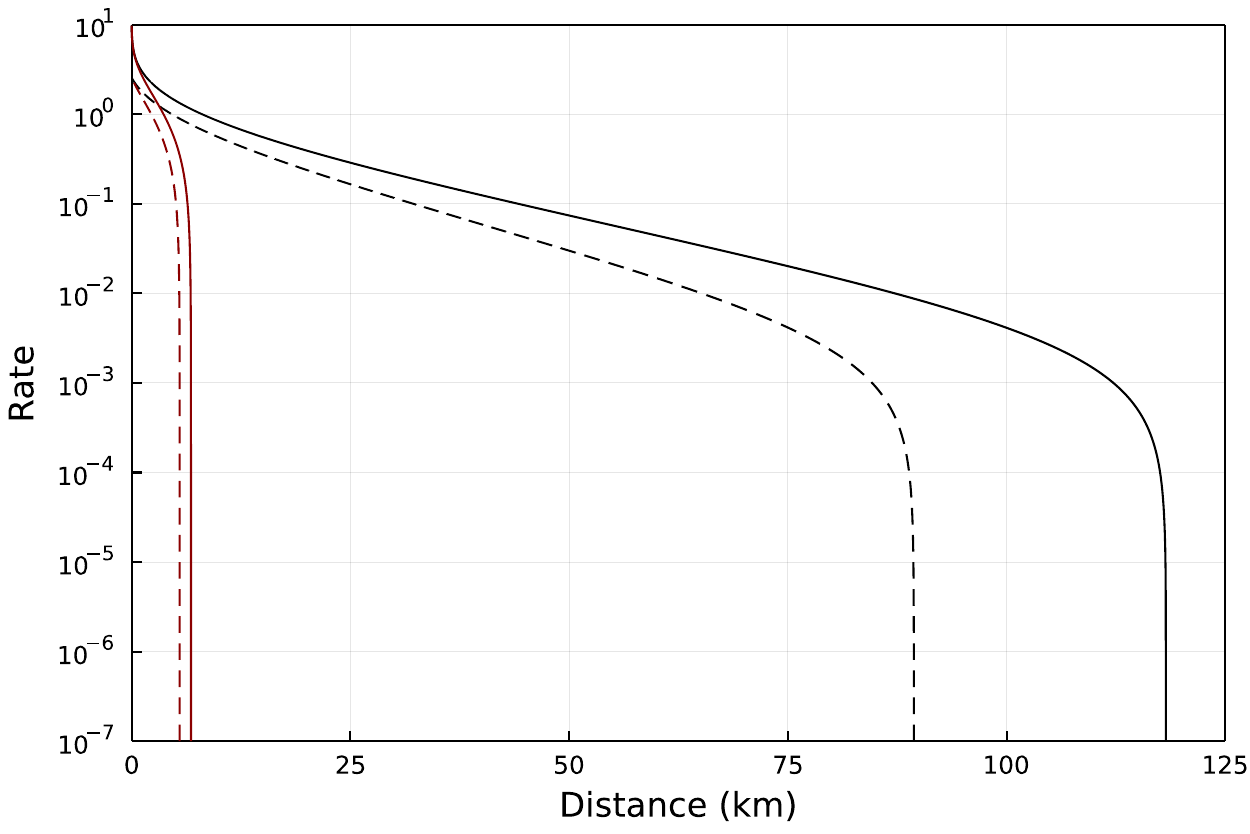}
    \caption{Asymptotic key rate (bits/use) of CV-MDI-QKD in the asymmetric configuration versus fibre distance ($0.2$dB/km) between Alice and Bob. Black lines refer to Alice at a fixed distance of $5$m from the relay, while Bob is at a variable fibre distance. The red lines represent the opposite scenario in which Bob's is fixed at $5$m from the relay and Alice's fibre distance is allowed to vary. Solid lines indicate the rates under pure loss and the dashed with thermal noise equal to $\bar{n}=0.05$. All rates are optimized with large modulation and perfect detection efficiency is assumed. \SP{ Figure adapted from Ref.~\cite{Fletcher:thesis2023}.}}
    \label{fig:2_user_assym_rate}
\end{figure}

In \cref{fig:2_user_assym_rate} we show the rate-distance scaling in the asymmetric configuration, where we assume that Alice is the encoder and Bob the decoder (this means that Bob needs to guess Alice's variable). If we fix the distance of Alice at $5$m from the relay, we can see that Bob can be very far. The distance scaling is much improved in this configuration, enabling communication at over $100$km between the two end parties. Indeed, it is shown in Ref.~\cite{Pirandola:CVMDI2015}, that if Bob's channel is affected by pure loss only, in the limit $\tau_A\rightarrow1$ (Alice brought arbitrarily close to the relay) that the rate only goes to zero for $\tau_B\rightarrow0$, which represents Bob arbitrarily far away. 

The reverse scenario in which Bob's position is fixed close to the relay is substantially less effective. Even with $\tau_B\rightarrow1$ and pure loss in Alice's channel, a positive rate can only be maintained to $6.8$km. This difference of performance is related to the distinction between direct and reverse reconciliation in CV QKD.  In the limit $\tau_A\rightarrow 1$ in which Alice (the encoder) approaches the relay, we recover a point-to-point no-switching protocol in reverse reconciliation. Conversely, with $\tau_B\rightarrow 1$ (decoder close to the relay), we have a point-to-point no-switching protocol in direct reconciliation.

\subsection{Post-selection}
\label{sec:postselection}
The distance of CV-MDI-QKD may be improved by introducing post-selection techniques. Post-selection was first utilized in QKD in Ref.~\cite{Silberhorn:3db2002} to overcome the apparent 3dB loss limit for CV QKD implied by beam-splitting attacks. Such technique exploits the fact that, even beyond 3dB, there are regions in the parameter space in which Alice and Bob's mutual information exceeds Eve's Holevo information and hence the rate is positive. By announcing the absolute values of the quadratures of their prepared coherent states, Alice and Bob can determine such regions in post-selection and keep only the most favourable instances of the protocol. They may then attempt reconciliation on the signs of the quadratures to produce a secure key. This technique improves the distance of CV QKD at the cost of reducing its rate at short ranges.

Ref.~\cite{Wilkinson:CVMDI2020} introduced post-selection for CV-MDI-QKD. In the prepare-and-measure formalism, Alice and Bob again prepare coherent states and each has access to zero-mean bi-variate Gaussian distributions, from which they draw two real variables for the values of the displacement of the $q$ and $p$ quadratures of the coherent states they will prepare. They encode the magnitude of their value for the $q$ displacement as $\mathbb{Q}_A$ and $\mathbb{Q}_B$ and the sign of the value as $\kappa_{A}$ and $\kappa_{B}$. Similarly, the absolute values drawn for the $p$ displacements are encoded as $\mathbb{P}_A$ and
$\mathbb{P}_B$, and the signs as $\kappa_{A}'$ and $\kappa_{B}'$. Thus, Alice and Bob prepare the coherent states $\ket{\alpha=\frac{1}{2}(\kappa_A\mathbb{Q}_A+i\kappa_A'\mathbb{P}_A)}$ and $\ket{\beta=\frac{1}{2}(\kappa_B\mathbb{Q}_B+i\kappa_B'\mathbb{P}_B)}$, respectively. 

These coherent states are sent to the relay as before and the measurement outcome $\gamma=\gamma_q+i\gamma_p$ is again announced publicly (where we have set $\gamma_q = q_{-}$ and $\gamma_p = p_{+}$). After the protocol ends, the parties undertake basis reconciliation to agree whether to use the $q$ or $p$ quadrature for each round of the protocol. If the $q$ quadrature is chosen, Alice and Bob reveal their values of $\mathbb{Q}_A$, $\mathbb{Q}_B$, $\kappa_A'\mathbb{P}_A$ and $\kappa_B'\mathbb{P}_B$ and will attempt to reconcile the signs $\kappa_A$ and $\kappa_B$. Alternatively if the $p$ quadrature is chosen, Alice and Bob reveal $\kappa_A\mathbb{Q}_A$, $\kappa_B\mathbb{Q}_B$, $\mathbb{P}_A$ and $\mathbb{P}_B$ and attempt to reconcile the signs $\kappa_A'$ and $\kappa_B'$.

Under phase-insensitive collective Gaussian attacks (such as the typical case of a collective attack based on two entangling cloners), the two quadratures are \GS{perturbed symmetrically and remain uncorrelated}. For this reason, we shall consider only the use of the $q$ quadrature. The measurement yields the distribution $p(\gamma_q|\kappa_A,\kappa_B,\mathbb{Q}_A,\mathbb{Q}_B)$ (note we have implicitly removed conditioning from variables on which there is no dependence) from which Bayes' theorem may be used to calculate $p(\kappa_A|\kappa_B,\mathbb{Q}_A,\mathbb{Q}_B,\gamma_q)$ and $p(\kappa_A,\kappa_B|\mathbb{Q}_A,\mathbb{Q}_B,\gamma_q)$ and hence determine the single point mutual information $\widetilde{I}(\kappa_A:\kappa_B|\mathbb{Q}_A,\mathbb{Q}_B,\gamma_q)$. Similarly we can the single point Holevo information $\widetilde{\chi}(E:\kappa_A|\mathbb{Q}_A,\mathbb{Q}_B,\gamma_q)$ and the single-point rate
\begin{align}
\widetilde{R}(\mathbb{Q}_A,\mathbb{Q}_B,\gamma_q) &=\widetilde{I}(\kappa_A:\kappa_B|\mathbb{Q}_A,\mathbb{Q}_B,\gamma_q) \nonumber \\ 
&-\widetilde{\chi}(E:\kappa_A|\mathbb{Q}_A,\mathbb{Q}_B,\gamma_q) ,
\end{align}
which is a rate conditioned on the values of the announced variables $\mathbb{Q}_A,\mathbb{Q}_B$ and the measurement outcome $\gamma_q$. By applying post-selection, Alice and Bob only select instances of the protocol when the single point rate is positive. Hence the overall postselected rate is given by:
\begin{equation}
\int \mathrm{max}\big[\widetilde{R}(\mathbb{Q}_A,\mathbb{Q}_B,\gamma_q),0\big] ~ p(\mathbb{Q}_A,\mathbb{Q}_B,\gamma_q) ~ d\mathbb{Q}~ d\gamma_q .
\end{equation}
Alternatively, denoting the region in parameter space in which the single point rate is positive as $\Gamma$ we may write the postselected rate as:
\begin{equation}\label{ps_asym_rate}
R=\int_{\Gamma} p(\mathbb{Q}_A,\mathbb{Q}_B,\gamma_q)~\widetilde{R}(\mathbb{Q}_A,\mathbb{Q}_B,\gamma) ~ d\mathbb{Q}~ d\gamma_q .
\end{equation}

\begin{figure}
    \centering
    \includegraphics[width=\linewidth]{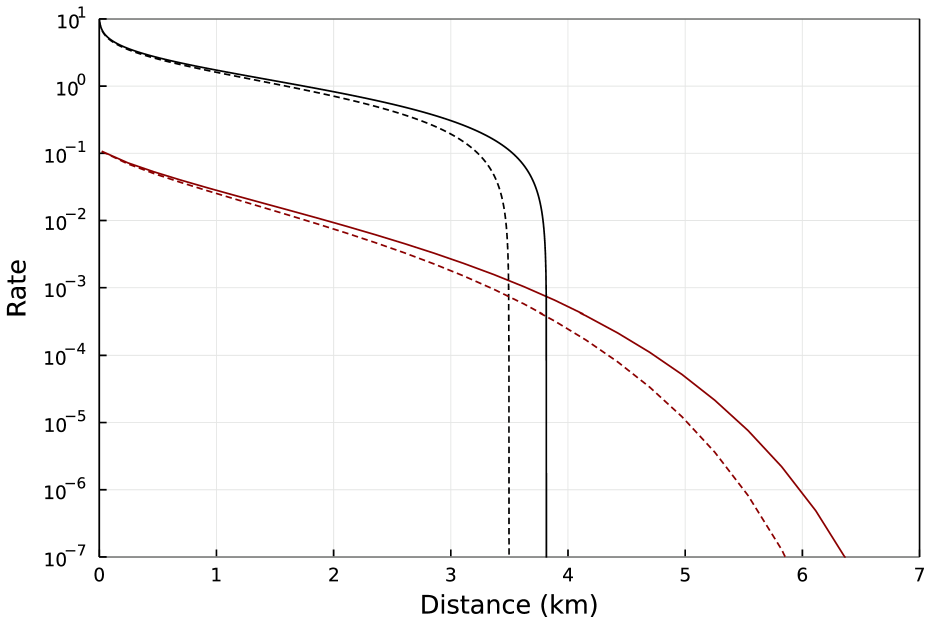}
    \caption{Comparative rate-distance performance of CV-MDI-QKD with and without post-selection in the symmetric configuration. The black lines indicate the original CV-MDI-QKD protocol \cite{Pirandola:CVMDI2015} and the red the equivalent protocol with post-selection~\cite{Wilkinson:CVMDI2020}. Solid lines are the rates under pure loss and dashed lines refer to thermal noise equal to $\bar{n}=0.05$ thermal photons. All the rates (bits/use) are optimized over the modulation of the prepared coherent states and assume perfect reconciliation and detection efficiency. \SP{ Figure adapted from Ref.~\cite{Fletcher:thesis2023}.}} \label{fig:two-party postsel rate}
\end{figure}

The rate-distance performance of the two equivalent protocols with and without post-selection are compared in \cref{fig:two-party postsel rate}. We see the trade-off between the improved long-distance performance of the post-selected protocol and the reduction in the rate at short distances more akin to DV-MDI-QKD. The post-selected protocol is more robust in the symmetric configuration, able to maintain a positive key rate out to ~6km.

\section{Composable finite-size security}
\label{sec:Comp&FS}

Composable finite-size security of CV-MDI-QKD was initially studied in Refs.~\cite{Lupo:FSCVMDI2018,Papanastasiou:CVMDI23} \GS{and later extended to free-space scenarios in Ref.~\cite{GhalaiiPRA2023}, including satellite implementations}. Finite-size means that we consider a finite number of signals exchanged in the protocol, which automatically leads to the use of estimators for evaluating channel parameters. Composability means that there is a security parameter associated with the protocol that quantifies its distance from an ideal implementation. This is called `epsilon' security and combines the failure probabilities of data-processing steps (PE, EC, and PA).

The most rigorous formula for the composable finite-size key rate of the basic, Gaussian-modulated version of CV-MDI-QKD can be derived using the tools developed in Ref.~\cite{Pirandola:QKDrates2024} for the security of general CV-QKD protocols. Here we use these tools to provide an improved formulation. 

Assume that Alice and Bob run a long session composed of $n_\mathrm{bks}$ blocks, each block having $N$ points. A point corresponds to a single protocol run, where the parties transmit a pair of coherent states to the relay and extract two correlated classical values from the process. We set $N=n+m$, where $m$ points of the block are used for PE while the remaining $n$ points contribute to key generation. Also assume that, in post-processing, the variables are digitalized into $d$ bits per letter. Then we have the following imperfections to consider:
\begin{itemize}
    \item Up to an error probability $\epsilon_{\mathrm{pe}}$, Alice and Bob construct $n_{\mathrm{pe}}$ worst-case estimators for the channel parameters (transmissivity and thermal noise of each link).

    \item Up to an error probability $\epsilon_{\mathrm{ent}}$, Alice and Bob estimate the entropy of the key.
    
    \item Up to a failure probability $1-p_{\mathrm{ec}}$, EC is successful and, up to an error probability $\epsilon_{\mathrm{cor}}$, called `epsilon correctness', the resulting corrected block is the same for Alice and Bob, so they have two identical string of bits.
    
    \item Up to an error probability $\epsilon_{\mathrm{sec}}$, called `epsilon secrecy', PA is successful. This means that Alice's and Bob's strings are compressed into shorter strings that are completely decoupled from Eve. In turn, one may decompose $\epsilon_{\mathrm{sec}}=\epsilon_{\mathrm{s}}+\epsilon_{\mathrm{h}}$, where $\epsilon_{\mathrm{s}}$ is known as smoothing parameter and $\epsilon_{\mathrm{h}}$ as hashing parameter.
    
    %\SP{In particular, $1-\epsilon_\text{s}$ is the probability the key will be uniformly distributed while $1-\epsilon_\text{h}$ is the probability the key will be completely uncorrelated from Eve.}

\end{itemize}

The protocol has total epsilon security $n_{\mathrm{bks}} \epsilon$, where
\begin{equation}
    \epsilon \le \epsilon_{\text{cor}}+\epsilon_{\text{s}}+\epsilon
_{\text{h}}+\epsilon_\text{ent}+ n_{\mathrm{pe}} \epsilon_{\text{pe}}.
\end{equation}
Assuming collective Gaussian attacks, the key rate is upper bounded by [see Eq.~(68) of~\cite{Pirandola:QKDrates2024}]
\begin{equation}
R \le \frac{p_{\text{ec}}[n \widehat{R}_{\infty}^{\text{pe}}  - n \delta_\text{ent}-\sqrt{n}\delta_\text{aep} + \theta]}{N}, \label{key_rate_2024}
\end{equation}
where 
\begin{align}
    &\delta_\text{ent}=\log_2(n)\sqrt{2 n^{-1} \ln(2/\epsilon_\text{ent})},\\
    &\delta_{\text{aep}}\simeq4\log_{2}\left(2^{d/2}+2\right)  \sqrt{\log_{2}(2/\epsilon_{\text{s}}^{2})},\\
    & \theta:=\log_2(2\epsilon^{2}_{\text{h}}\epsilon_{\text{cor}}),
\end{align}
and $\widehat{R}_{\infty}^{\text{pe}}$ is the asymptotic key rate computed from PE. 

It is in the specific expression of $\widehat{R}_{\infty}^{\text{pe}}$ that enters the features of the CV-MDI-QKD protocol. In particular, this is given by
\begin{equation}
\widehat{R}_\infty^{\text{pe}}=\widehat{\xi}\widehat{I}_{AB}-[I_E]_\text{wc},
\end{equation}
where the various terms depend on estimators and worst-case estimators of $n_{\mathrm{pe}}=3$ parameters: the link transmissivities, $\tau_A$ and $\tau_B$, and the total excess noise~\cite[Eq.~(10)]{Papanastasiou:CVMDI23}.
%\SP{[*CITE*]}
\GS{For an uncorrelated two-mode attack with two independent entangling cloners ($g'=g=0$), the total excess noise takes the form
% \begin{align}
% \Xi=&\frac{1}{2}\Big ((1-\tau_A)(\omega_A-1)+(1-\tau_B)(\omega_B-1)\Big) \notag \\&+g\sqrt{(1-\tau_A)(1-\tau_B)}.
% \end{align}
\begin{align}
\Xi= \frac{1}{2}\Big (\tau_A \zeta_A+\tau_B \zeta_B\Big), 
\end{align}
where $\zeta_k:=(1-\tau_k)(\omega_k-1)/\tau$ is the excess noise of an individual link $k=A$ (Alice-relay) or $B$ (Bob-relay).}

The mutual information $\widehat{I}_{AB}$ is calculated from the estimators  $\widehat{\tau}_A$, $\widehat{\tau}_B$, and $\widehat{\Xi}$, while Eve's Holevo information $[I_E]_\text{wc}$ is computed from the worst-case estimators $\tau_A^\text{wc}$, $\tau_B^\text{wc}$, and $\Xi_\text{wc}$. In the presence of a stable configuration, these values can be computed over $n_{\mathrm{bks}} m$ points (see Ref.~\cite{Papanastasiou:CVMDI23} for more details on their explicit expressions).
%\SP{(see discussion around~\cite[Eq.~(63)]{Papanastasiou:CVMDI23}for more details on their explicit expressions).} 
In addition, the error reconciliation efficiency is given by
\begin{equation}
\widehat{\xi}=\frac{\widehat{H}_{\mathrm{key}}-n^{-1}\text{leak}_\text{ec}}{\widehat{I}_{AB}},
\end{equation}
where $\widehat{H}_{\mathrm{key}}$ is the estimated Shannon entropy of the raw key (see, for example, ~\cite[Eqs.(46) and~(47)]{Mountogiannakis:CompFS2022}) and leak$_\text{ec}$  upper-bounds the bits of information leaked during EC. 

\begin{figure}[h]
    \centering \includegraphics[width=0.99\linewidth]{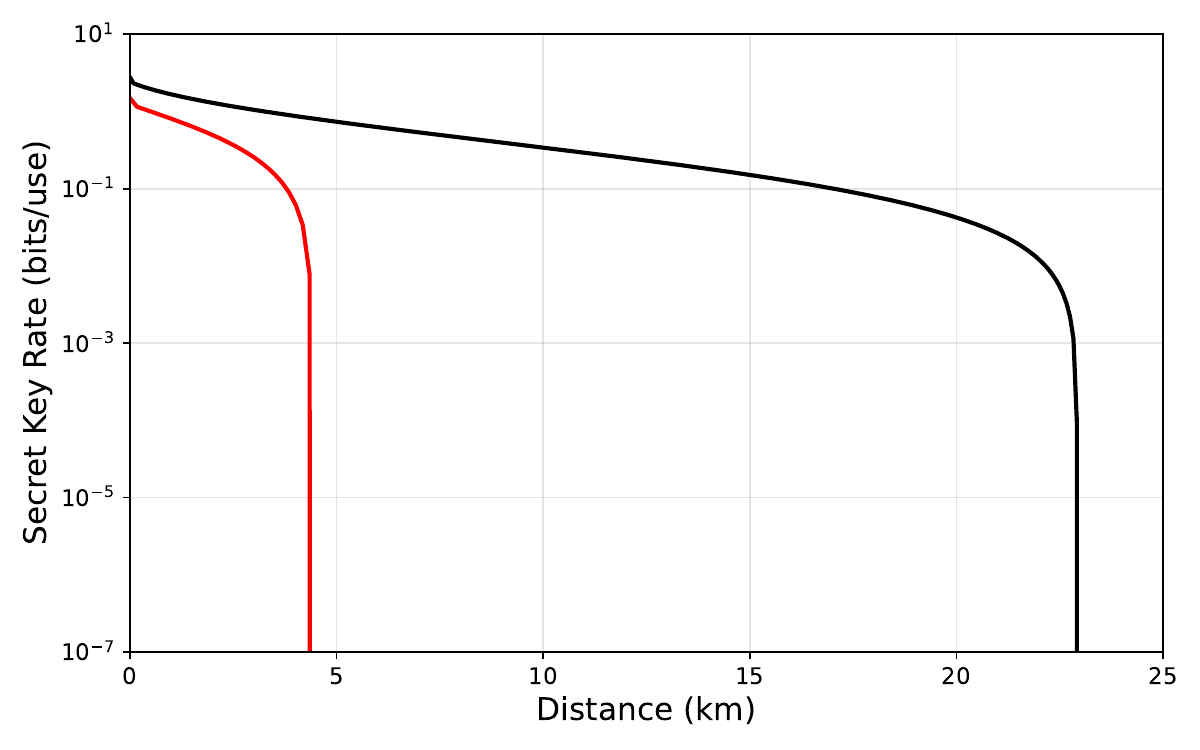}
    \caption{\GS{Composable-secure secret key rate (bits/use) of the symmetric (red line) and asymmetric (black line)  CV-MDI-QKD protocol versus fibre distance with standard loss-rate $0.2$dB/km (the distance in the plot is the sum of the fibre distance of the two links). The distance of Alice from the relay is $5$m for the asymmetric configuration. The values of the excess noise for the two individual links are $\zeta_A=\zeta_B=0.01$ and we assume the two-mode uncorrelated attack ($g'=g=0$). We have set $\widehat{\xi}=0.98$, $p_\text{ec}=0.95$, $d=14$ and a block size of $N=10^7$. The electronic noise and detection efficiency of the relay are set to $u_\text{el}=0.01$ and $\eta_\text{eff}=0.98$. All epsilon parameters are individually set to $2^{-32}$, while the PE ratio and the modulation variance are optimized. See Ref.~\cite{Papanastasiou:CVMDI23} for details about all the parameters.}} 
    \label{fig:comp_rate}
\end{figure}

\GS{In Fig.~\ref{fig:comp_rate}, we show the composable-secure key rate of the CV-MDI-QKD protocol in the symmetric and asymmetric configurations. We can see how the requirement for composable security takes a toll in terms of achievable distance by comparing the results with those in Fig.~\ref{fig:2_user_assym_rate}. For example, in the asymmetric configuration, the asymptotic key rate stretches up to about $100$~km, while this distance is reduced to about $25$~km for the composable key rate.}

\section{Network Complexity}
\label{sec:NetComplex}

\GS{In this section, we discuss how CV-MDI-QKD can be extended to network scenarios. First, we consider an extension to a star network with an arbitrary number of users. We then consider the specific case of a 3-node star network, for which we introduce the technique of post-selection, with the aim of improving the distribution distance from the central relay.}

\subsection{\GS{CV-MDI-QKD star network}}
\label{sec:star}
\GS{Let us} consider how CV-MDI-QKD can be extended from the bipartite case to $N$ users. Since MDI-QKD invariably requires the use of a central untrusted relay, it is natural to consider a star network topology in which each of the $N$ users connects to a single central relay. This approach is made possible by means of a generalized Bell detection at the central relay which creates shared correlations that can be post-processed into a secure key common to all parties (`key conferencing'). Overall, a CV-MDI-QKD network is a specific example of a Gaussian quantum network with untrusted relays, as generally modeled in Ref.~\cite{Ghalaii:GQNets2022}.

% As before, we consider the protocol in the entanglement based representation. Each user sends the $A$ mode of identical TMSVs (as given in \cref{eq:TMSV}) to the central detector relay. The optimal attack strategy is equivalent to passing each incoming mode through a thermal channel, with transmissivity $\eta$ and thermal noise $\bar{n}$
% which results in all of the end users' states arriving at the relay having the following covariance matrix:

% \begin{equation}
% \boldsymbol{V}^{(i)}_{AB}=\begin{pmatrix}
% x \boldsymbol{I} & z\boldsymbol{Z} \\
%  z\boldsymbol{Z} & y \boldsymbol{I}
% \end{pmatrix} 
% \end{equation}
% where the index $i$ indicates the end user; $x=\eta_i\mu+(1-\eta_i)(2\bar{n}+1)$, $y=\mu$ and $z=\sqrt{\eta_i(\mu^2-1)}$.

\begin{figure}[h]
    \centering
    \includegraphics[width=0.5\linewidth]{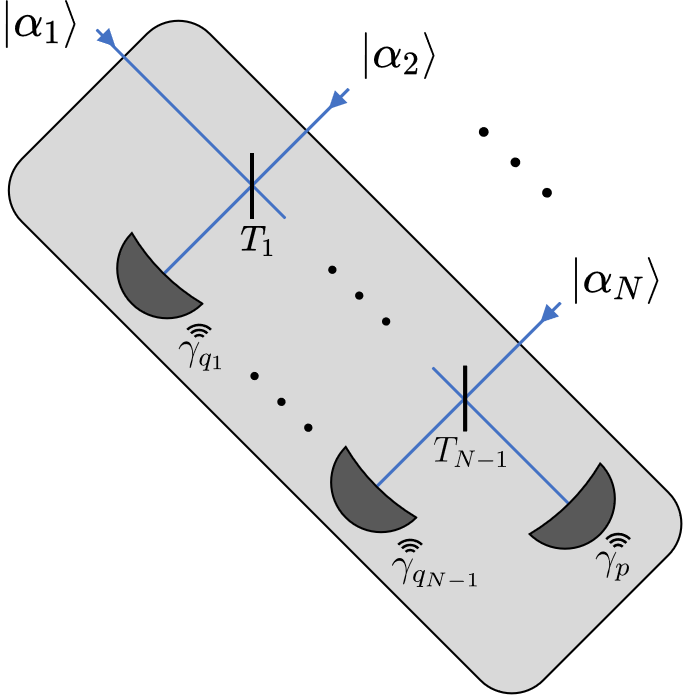}
    \caption{Structure of the detector relay for CV-MDI-QKD conferencing. \SP{ Figure adapted from Ref.~\cite{Ottaviani:QConf2019}.}}
    \label{fig:star detector}
\end{figure}

The structure of the relay for CV-MDI-QKD is depicted in Fig.~\ref{fig:star detector}. Each party prepares coherent states $\ket{\alpha_i}$ which are sent through thermal-loss channels to the detector relay. The detector consists of a cascade of $N-1$ beamsplitters of increasing transmissivity. Each output mode on the left of the diagram undergoes homodyne detection in the $q$ quadrature and the final mode at the bottom is homodyned in the $p$ quadrature. The results of the measurements are announced publicly.

% The structure of the detector is depicted in Fig. \ref{fig:star detector} and its effect in the entanglement based representation is to project the overall state of the end users $B$ modes into a Gaussian state, completely symmetric between all the $B$ modes with GHZ-like correlations.
% \begin{equation}
% \boldsymbol{V}_{\boldsymbol{B}|\gamma}=\begin{pmatrix}
% \boldsymbol{\Delta} & \boldsymbol{\Gamma} & \cdots & \boldsymbol{\Gamma} \\
% \boldsymbol{\Gamma} & \boldsymbol{\Delta} & \ddots & \vdots \\
% \vdots & \ddots & \ddots & \boldsymbol{\Gamma} \\
% \boldsymbol{\Gamma} & \cdots & \boldsymbol{\Gamma} & \boldsymbol{\Delta}
% \end{pmatrix}
% \label{eq:covar mdi star}
% \end{equation}

% $\boldsymbol{\Gamma}=\text{diag}\big(\frac{z^2}{Nx} ,-\frac{z^2}{Nx}\big)$ 
% $\boldsymbol{\Delta}=\text{diag}\big(y-\frac{(N-1)z^2}{Nx},y-\frac{z^2}{Nx}\big)$ 
% from which we may observe that the correlations between each of the $B$ modes scales as $1/N$. In the two user case eq. \ref{eq:covar mdi star} is equivalent to the state generated between the two users $B$ modes. In this case the relative entropy of entanglement of the state $\hat{\rho}_{\small{B_1,B_2|\gamma}}$ upper bounds the rate of the protocol so that the rate is only positive when the user's $B$ modes are entangled. With $N>2$ users, the rate may remain positive even when the reduced bipartite state between two users $B$ modes is not entangled, since the other users states are effectively act as trusted noise.

\begin{figure}[h]
    \centering
    \includegraphics[width=\linewidth]{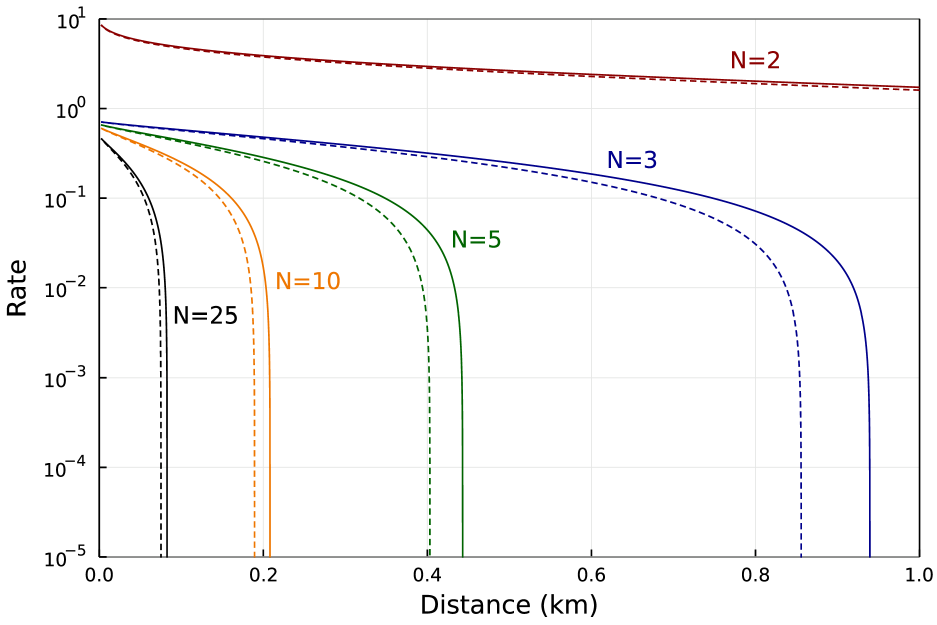}
    \caption{Rate (bits/use) of a CV-MDI-QKD star network in the symmetric configuration as a function of the fibre distance of each user from the central relay. The variances of the Gaussian distributions used to modulate the coherent states are selected to optimize the maximum distance at which a positive secret key rate can be maintained. Solid lines represent the rate under pure loss ($\bar{n}=0$) and the dashed lines under non-zero thermal noise ($\bar{n}=0.05$). \SP{ Figure adapted from Ref.~\cite{Fletcher:thesis2023}.}}
    \label{fig:Coherent Star Rates}
\end{figure}

The rates achieved by the star network module for various numbers of end users and amounts of thermal noise are shown in Fig.~\ref{fig:Coherent Star Rates}. The users are all equidistant to the central relay (symmetric configuration) and the rates shown are optimized for the maximum distance at which they maintain a positive key rate. There is a fundamental trade-off between the number of users and the maximum distance at which a positive key rate may be maintained. More precisely, from Fig.~3b of Ref.~\cite{Ottaviani:QConf2019}, we see a $\simeq 2/N$ scaling for the maximum radius in km of the star network in terms of the number of users $N$.

Due to the limitations of linear optics, it is important to remark that similar designs for multi-partite MDI-QKD cannot be implemented with discrete variables. A detailed discussion may be found in the supplementary information of Ref.~\cite{Ottaviani:QConf2019}. Finally, it is also worth to mention that a squeezed version of multi-partite CV-MDI-QKD can achieve higher rates than the standard version with coherent states. \GS{As discussed in Ref.~\cite[Supp. Note~4]{Ottaviani:QConf2019}, higher rates may be achieved if the parties perform the protocol in an entanglement-based configuration where they use TMSV states, sending one mode to the relay while retaining the other. After the generalized Bell detection is performed, they would then apply conditional local squeezing operations to their kept modes before measuring them with heterodyne detection.}

\subsection{CV-MDI-QKD 3-node star network with post-selection}
\begin{figure}[h]
    \centering
    \includegraphics[width=0.58\linewidth]{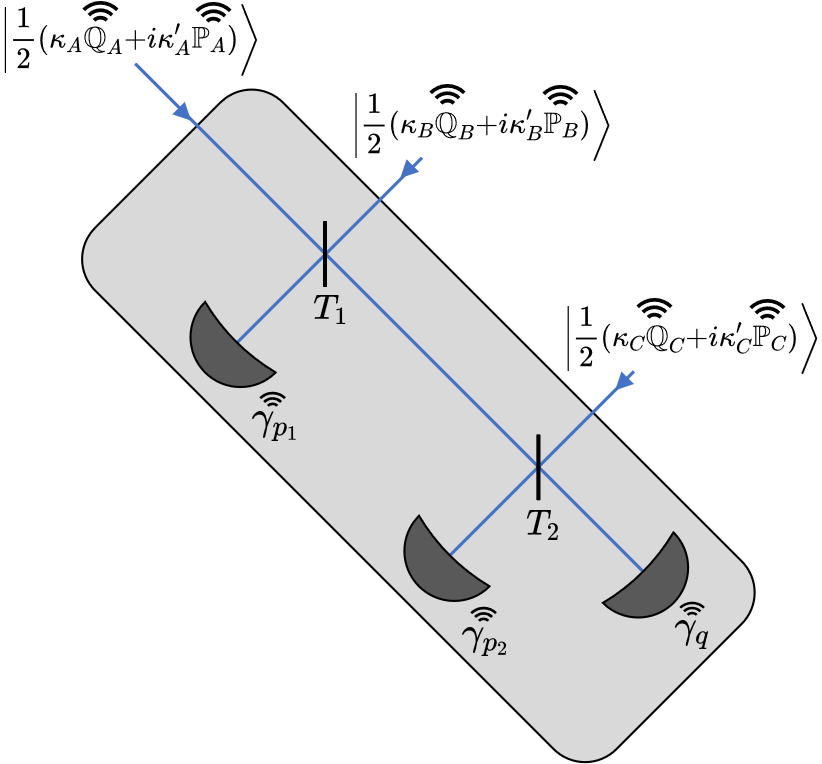}
    \caption{3-user CV-MDI-QKD star network in post-selection. Two beamsplitters mix input modes with transmissivities $T_1=\frac{1}{2}$ and $T_2=\frac{2}{3}$. The states undergo two $p$ homodyne detections and one $q$ homodyne detection. In this case, the parties will attempt reconciliation between $\kappa_A,\kappa_B,\kappa_C$. The other orientation is physically equivalent to Fig. \ref{fig:star detector}, in which the parties attempt reconciliation on $\kappa_A',\kappa_B',\kappa_C'$. In both cases at the end of the protocol all the end users publicly broadcast the magnitudes of the quadratures of their prepared states. Figure adapted from Ref.~\cite{Fletcher:3nodeCVMDI2022}.}
    \label{fig:3node reverse config}
\end{figure}

As with the two user case, post-selection techniques may be applied to such a setup to improve the long distance performance of the star network module. In particular, the 3 end user case has been recently investigated \cite{Fletcher:3nodeCVMDI2022}. In this protocol the 3 parties prepare states of the form $\ket{\frac{1}{2}(\kappa_A\mathbb{Q}_i+i\kappa_i'\mathbb{P}_i)}$ again having drawn $q_i$ and $p_i$ from a zero mean Gaussian and encoding the signs as $\kappa_i$ and $\kappa_i'$ and the magnitudes as $\mathbb{Q}_i$ and  $\mathbb{P}_i$. The structure of the detector is the same as the star network described in \cref{sec:star} with one key change: The detector is randomly switched between performing $N-1 ~ q$ homodyne detections and $1 ~ p$ homodyne detection and $N-1 ~ p$ homodyne detections and $1 ~ q$ homodyne detection. This second configuration is shown in the schematic in \cref{fig:3node reverse config}. The chosen configuration is publicly announced after the detections.

\begin{figure}[h]
    \centering
    \includegraphics[width=\linewidth]{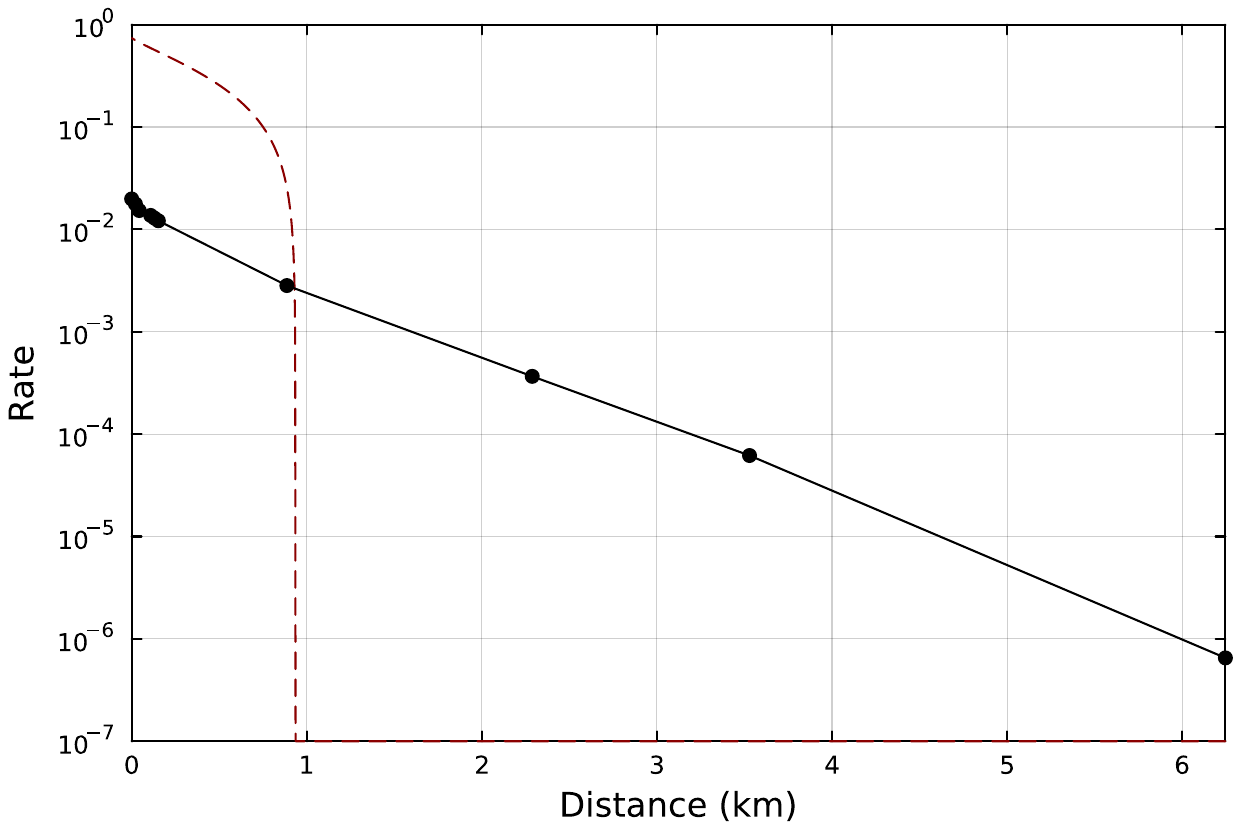}
    \caption{Rate of the 3-user CV-MDI-QKD network in post-selection assuming the symmetric configuration (all parties equidistant from the central relay). The key rate (bits/use) is plotted versus fibre distance of each party assuming perfect detector efficiency and unit variance for the Gaussian modulations. For comparison, we show the rate of an equivalent 3-party star network without post-selection~\cite{Ottaviani:QConf2019} and optimized parameters (red dashed line). In this plot, the links are all pure-loss channels. Figure adapted from Ref.~\cite{Fletcher:3nodeCVMDI2022}.}
    \label{fig:3-party postselection rate}
\end{figure}

As in \cref{sec:postselection} the postselected rate may be computed by performing an integral over the region $\Gamma$ where the single point rate $\widetilde{R}$ is positive:
\begin{equation}
R=\int_{\Gamma} p(\pmb{\mathbb{Q}},\gamma_q) \ \widetilde{R}(\pmb{\mathbb{Q}},\gamma_p)  \ d\pmb{\mathbb{Q}}~ d\gamma_q  
\label{eq:3partypostint}
\end{equation}
where we have collected $\pmb{\mathbb{Q}}=(\mathbb{Q}_A,\mathbb{Q}_B,\mathbb{Q}_C)$.
The results are shown in Fig.~\ref{fig:3-party postselection rate}, comparing the postselected and non-postselected protocols under pure loss. We see the same reduction in the rate at short distances but improved long-distance performance. The numerical difficulty in performing the high dimensional integral in \cref{eq:3partypostint} limits the analysis to pure loss attacks but the work provides proof of principle that such post-selection techniques may be applied to multi-party CV-MDI-QKD conferencing.

\section{Experimental CV-MDI-QKD}
\label{sec:experiments}
CV-MDI-QKD enhances security by eliminating side-channel vulnerabilities on detection. This security benefit comes at the expense of increased implementation complexity compared to one-way CV-QKD. Consequently, despite significant progress in theory, only a few experimental demonstrations of CV-MDI-QKD have been reported, \GS{as summarized in Table~\ref{tab:table1}.}

\begin{figure*}
\vspace{+.1cm}
\includegraphics[width=\textwidth]{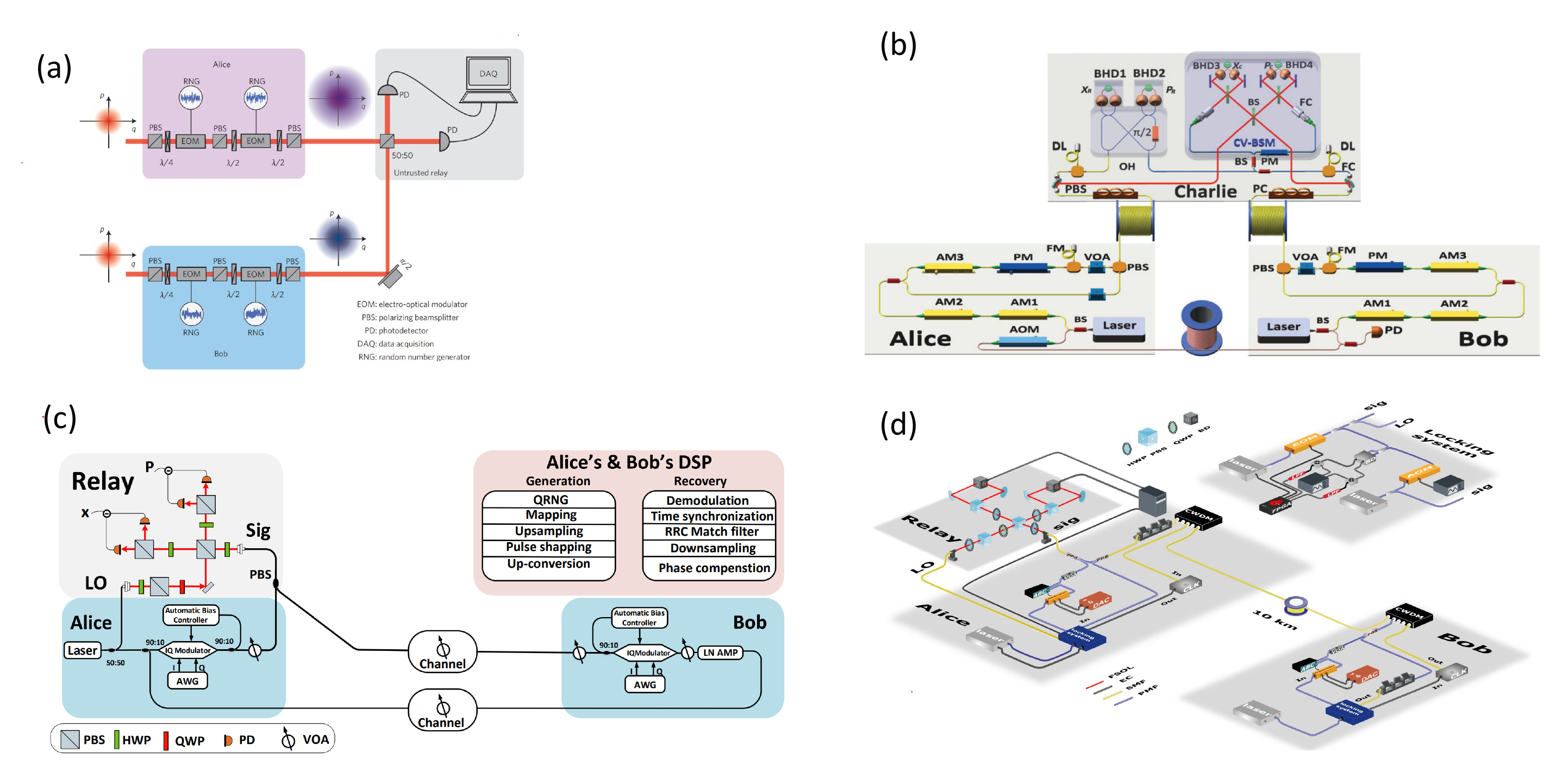}
\caption{ Experimental CV-MDI-QKD based on Gaussian-modulated coherent states. (a) ~\cite{Pirandola:CVMDI2015} (b) ~\cite{tian2022experimental} (c) ~\cite{hajomer2022high} (d)~\cite{hajomer2023real}. BS: beam splitter; AM: amplitude modulator; AOM: acousto-optic modulator; PM: phase modulator; FM: Faraday mirror; VOA: variable optical attenuator; PBS: polarizing beam splitter; PD: photodetector; PC: polarization controller; FC, fibre collimator; DL: delay line; AWG: arbitrary waveform generator; HWP: Half-wave plate; QWP: quarter-wave plate; BHD: balanced homodyne detector; DAC: Digital-to-analog converter; ABC: Automatic bias controller; CWDM: Coarse Wavelength division multiplexing; FPGA: Field-programmable gate array. Figures adapted from Refs.~\cite{Pirandola:CVMDI2015},~\cite{tian2022experimental},~\cite{hajomer2022high}, and~\cite{hajomer2023real}. 
} 
\label{fig:ExpDemo}
\end{figure*}

In 2015, the earliest proof-of-concept demonstration of CV-MDI-QKD was performed using a free-space optical setup, as illustrated in Fig.~\ref{fig:ExpDemo}(a) \cite{Pirandola:CVMDI2015}. In this experiment, coherent states were generated by modulating a continuous-wave laser at 1064~nm with amplitude and phase modulators driven by Gaussian noise. Phase synchronization between Alice's and Bob's signals was achieved using piezoelectric transducers. The CV Bell measurement was implemented using a balanced beam splitter combined with homodyne detectors measuring orthogonal quadratures. The setup achieved a secret key rate of 0.1 bits per relay use under emulated channel losses of 2\% for Alice and 60\% for Bob, assuming a reconciliation efficiency of 97\%. While this work showed the fundamental viability of CV-MDI-QKD, it did not incorporate essential practical features like fibre-optic transmission, independent laser sources at Alice's and Bob's stations operating at telecommunication wavelengths, and clock synchronization between the transmitters, making it unsuitable for practical deployment.

\begin{table*}[t]
   \centering
\GS{\caption{\textbf{Comparison of Experimental Parameters in Recent CV-MDI-QKD Demonstrations}} \label{tab:table1}
%\resizebox{\hsize}{!}{
\begin{tabular}{|c|c|c|c|c|c|c|}
%         \hline  \multicolumn{3}{|c|}{\textbf{Entry}}  \\
         \hline \bf Ref.  &\bf Configuration & \bf channel type & \bf loss& \bf repetition rate  & \bf key rate, bits/relay use  &\bf Security \\
         \hline  \cite{Pirandola:CVMDI2015} & asymmetric & Free space & 4 dB & 100 kHz& 0.1& Asymptotic\\
         \hline  \cite{tian2022experimental} & asymmetric & Fiber & 2.02 dB & 500 kHz & 0.19 & Asymptotic\\
         \hline  \cite{hajomer2022high} & asymmetric & VOA & 2 dB & 5 MHz & 0.12 & Asymptotic\\
         \hline  \cite{hajomer2023real} & asymmetric & Fiber & 2.5 dB & 20 MHz & 0.13 & Finite-size\\
         \hline
\end{tabular}
}
\end{table*}

A more practical CV-MDI-QKD implementation was demonstrated in 2022~\cite{tian2022experimental}, as illustrated in Fig.~\ref{fig:ExpDemo}(b). This system employed a 10 km standard single-mode fibre (SMF) channel and independent CW lasers operating at the telecom wavelength of 1550 nm. The two lasers were frequency-locked by sending part of Alice's laser through a fibre channel to Bob's station, which interfered with a portion of  Bob's laser. The interference result was then fed as an error signal to the phase-locked loop. The system used a 50 ns pulse carved at a repetition rate of 500 kHz. The pulses were split into quantum pulses modulated by cascaded amplitude and phase modulators, pilot pulses used as LO (coming from Bob's station) and phase-reference pulses (coming from Alice's station) to compensate for the fast phase drift between the LO and Alice's signal. The quantum and pilot pulses were time and polarization multiplexed at Alice's and Bob's stations using a combination of polarizing beam splitter, delay line, and Faraday mirror before transmitting them through SMF. At the relay, the quantum and the pilot pulses were first demultiplexed. To estimate the fast phase drift, heterodyne detection was performed by tapping off part of the LO pulses and interfering them with the reference pulses in a 90-degree optical hybrid measuring the amplitude quadrature $X_R$ and the phase quadrature $P_R$. The measurement outcomes were then used to estimate the fast phase drift for each signal pulse as $\Delta\theta = \tan^{-1}(P_R / X_R)$. As for the slow phase drift estimation, Alice and Bob used another set of phase-calibration pulses, which were different from the phase-reference pulses. After estimating the total phase drift, Alice and Bob applied quadrature remapping to their data. With this frequency and phase locking system, a low excess noise of 0.0045 SNU was achieved, enabling a key rate of 0.19 bit per relay use at an error correction efficiency of 97\%. This work demonstrated the feasibility of building a CV-MDI-QKD system over fibre channels. However, the complexity of the experiment is rather high. For instance, a separate heterodyne receiver is required to implement the complex frequency and phase locking system, and because it is a pulsed system, an additional amplitude modulator is required for pulse carving. The combination of polarization and time multiplexing also adds another fold of complexity. Moreover, the experiment does not take into account how to synchronize distant transmitters. 

The development of the next generation of CV-MDI-QKD systems focuses on simplifying the system structure and increasing the repetition rate. Figure~\ref{fig:ExpDemo}(c) shows a recent experiment towards this goal~\cite{hajomer2022high}. In this experiment, the frequency and phase locking system, as well as pulse carving were removed. This is done by sharing a 1550 nm CW laser between the communication parties with a low noise amplifier (LN AMP) deployed at Bob’s station to avoid frequency locking. Besides, the pulse carving was replaced by digital pulse shaping using a root-raised cosine filter. This allowed the transmitters to operate at a symbol rate of 5 MBaud, which is one order of magnitude larger than the previous demonstration ~\cite{Pirandola:CVMDI2015,tian2022experimental}. Furthermore, the CV Bell detection was realized without phase locking utilizing a new relay structure leveraging the concept of a polarization-based 90-degree hybrid. The quantum efficiency of the relay was 94\%. Combining this structure with digital signal processing (DSP), the propagation delay and phase drift were estimated using reference symbols. The system generated keys at a rate of 0.12 bit per relay use over channel loss of 2 dB, assuming an information reconciliation efficiency of 97\% .   

The coexistence of CV-MDI-QKD with classical communication channels is pivotal for real-world adoption. A recent system design has addressed this issue, taking into account practical aspects such as clock synchronization and the utilization of SMF channels~\cite{hajomer2023real}. In this system, an asymmetric network configuration was employed, with the relay being co-located at Alice's station, as shown in Fig.~\ref{fig:ExpDemo}(D). Both senders, Alice and Bob, utilized the same DSP module for quantum state preparation, following the approach in Ref.~\cite{hajomer2022high}. To achieve synchronization between Alice's and Bob's stations, a 10 MHz reference clock was utilized at Bob's station to generate an optical clock at 1310~nm, which co-propagated with the quantum signal to Alice's station. Furthermore, real-time phase locking was integrated into the system to lock two independent lasers, eliminating the need for an additional fibre channel as described in Ref.~\cite{tian2022experimental}.  This significantly simplifies the system structure, making it more suitable for practical adoption. The system operated at a symbol rate of 20 Mbaud and demonstrated key generation in finite-size regimes. This resulted in a two-order-of-magnitude improvement in the key rate compared to previous demonstrations~\cite{tian2022experimental, Pirandola:CVMDI2015}. 

\GS{Despite these advances, further improvements are possible. For instance, enhanced DSP techniques could eliminate the need for explicit laser phase locking and clock synchronization. However, the achieved symbol rates in CV-MDI-QKD still lag behind the state-of-the-art in prepare-and-measure coherent-state QKD systems~\cite{hajomer2024continuous}. This limitation arises primarily from two factors: (1) bandwidth constraints in the high-efficiency relay, particularly in the readout electronics of balanced homodyne detectors, which often rely on discrete components with parasitic limitations that restrict both speed and noise performance; and (2) the stringent timing requirements at the relay beam splitter. Unlike one-way protocols, CV-MDI-QKD requires precise temporal overlap of optical pulses from independent sources. As the symbol rate increases and pulse durations shorten, synchronization becomes more challenging, directly impacting the interference visibility and system performance.}

\section{Conclusions}\label{sec:conclusions}

CV-MDI-QKD amalgamates two groundbreaking concepts within QKD: the techniques of CV-QKD and MDI-QKD. On the one hand, this approach streamlines quantum hardware, enabling the use of non-single-level coherent-state sources and room-temperature homodyne or heterodyne detectors. On the other, it establishes single-hop end-to-end encryption, where Alice and Bob extract a secret key even when the middle party is untrusted, i.e., potentially operated by an eavesdropper. In this short review, we have discussed the features of this protocol, its state-of-art in terms of security proofs, and experimental implementations. 

While the promise of CV-MDI-QKD is very appealing for the entire field of quantum cryptography, its robustness to practical loss affects its performance in the symmetric configuration where Alice and Bob are approximately equidistant from the central relay. Besides the early development of the asymmetric version of the protocol, efforts have been directed to develop variants, e.g., based on post-selection, that would allow to mitigate this fragility and reach better distances when operated symmetrically. 

The overall scope of this short review is to boost further research in this area. The main theoretical goals are: (i) develop a symmetric variant able to further increase the distance of both parties from the central relay; and (ii) develop a multi-partite network variant for quantum conferencing able to mitigate the $O(1/N)$ radius scaling with respect to the number of users $N$. From an experimental perspective there are several avenues for exploration: (i) developing a simpler system that is hardware-agnostic and phase-locking free; (ii) increasing the system's repetition rate; and (iii) integrating photonics and electronics components of the relay and transmitters.  

\GS{CV-MDI-QKD presents a promising solution for securing high-rate local area networks (LANs) where multiple devices communicate through a central, potentially untrusted relay. The compatibility with standard telecom components and the potential for high key generation rates over short distances make it especially suitable for densely connected LAN environments, such as data centres or enterprise networks, where secure and efficient key exchange is critical.}

\vspace{+0.2cm}

\textbf{Acknowledgements}. AH and TG acknowledge support from the Innovation Foundation Grand Solutions project CyberQ (grant agreement~3200-00035B), the Danish National Research Foundation, Center for Macroscopic Quantum States (bigQ, DNRF142), and the Carlsberg Foundation (grant agreement number~CF21-0466). This project has also received funding from the European Union’s Horizon Europe research and innovation programme under the project ``Quantum Security Networks Partnership'' (QSNP, grant agreement No.~101114043). 

\bibliography{references}

\begin{thebibliography}{102}%
\makeatletter
\providecommand \@ifxundefined [1]{%
 \@ifx{#1\undefined}
}%
\providecommand \@ifnum [1]{%
 \ifnum #1\expandafter \@firstoftwo
 \else \expandafter \@secondoftwo
 \fi
}%
\providecommand \@ifx [1]{%
 \ifx #1\expandafter \@firstoftwo
 \else \expandafter \@secondoftwo
 \fi
}%
\providecommand \natexlab [1]{#1}%
\providecommand \enquote  [1]{``#1''}%
\providecommand \bibnamefont  [1]{#1}%
\providecommand \bibfnamefont [1]{#1}%
\providecommand \citenamefont [1]{#1}%
\providecommand \href@noop [0]{\@secondoftwo}%
\providecommand \href [0]{\begingroup \@sanitize@url \@href}%
\providecommand \@href[1]{\@@startlink{#1}\@@href}%
\providecommand \@@href[1]{\endgroup#1\@@endlink}%
\providecommand \@sanitize@url [0]{\catcode `\\12\catcode `\$12\catcode
  `\&12\catcode `\#12\catcode `\^12\catcode `\_12\catcode `\%12\relax}%
\providecommand \@@startlink[1]{}%
\providecommand \@@endlink[0]{}%
\providecommand \url  [0]{\begingroup\@sanitize@url \@url }%
\providecommand \@url [1]{\endgroup\@href {#1}{\urlprefix }}%
\providecommand \urlprefix  [0]{URL }%
\providecommand \Eprint [0]{\href }%
\providecommand \doibase [0]{http://dx.doi.org/}%
\providecommand \selectlanguage [0]{\@gobble}%
\providecommand \bibinfo  [0]{\@secondoftwo}%
\providecommand \bibfield  [0]{\@secondoftwo}%
\providecommand \translation [1]{[#1]}%
\providecommand \BibitemOpen [0]{}%
\providecommand \bibitemStop [0]{}%
\providecommand \bibitemNoStop [0]{.\EOS\space}%
\providecommand \EOS [0]{\spacefactor3000\relax}%
\providecommand \BibitemShut  [1]{\csname bibitem#1\endcsname}%
\let\auto@bib@innerbib\@empty
%</preamble>
\bibitem [{\citenamefont {Bennett}(1984)}]{BB84}%
  \BibitemOpen
  \bibfield  {author} {\bibinfo {author} {\bibfnamefont {C.~H.}\ \bibnamefont
  {Bennett}},\ }\bibfield  {booktitle} {\emph {\bibinfo {booktitle}
  {Proceedings of IEEE International Conference on Computers Systems and Signal
  Processing}},\ }\href@noop {} {\ ,\ \bibinfo {pages} {175} (\bibinfo {year}
  {1984})}\BibitemShut {NoStop}%
\bibitem [{\citenamefont {Bennett}\ and\ \citenamefont
  {Brassard}(2014)}]{BB84:2004}%
  \BibitemOpen
  \bibfield  {author} {\bibinfo {author} {\bibfnamefont {C.~H.}\ \bibnamefont
  {Bennett}}\ and\ \bibinfo {author} {\bibfnamefont {G.}~\bibnamefont
  {Brassard}},\ }\href {https://doi.org/10.1016/j.tcs.2014.05.025} {\bibfield
  {journal} {\bibinfo  {journal} {Theor. Comput. Sci.}\ }\textbf {\bibinfo
  {volume} {560}},\ \bibinfo {pages} {7} (\bibinfo {year} {2014})}\BibitemShut
  {NoStop}%
\bibitem [{\citenamefont {Pirandola}\ \emph {et~al.}(2020)\citenamefont
  {Pirandola}, \citenamefont {Andersen}, \citenamefont {Banchi}, \citenamefont
  {Berta}, \citenamefont {Bunandar}, \citenamefont {Colbeck}, \citenamefont
  {Englund}, \citenamefont {Gehring}, \citenamefont {Lupo}, \citenamefont
  {Ottaviani}, \citenamefont {Pereira}, \citenamefont {Razavi}, \citenamefont
  {Shaari}, \citenamefont {Tomamichel}, \citenamefont {Usenko}, \citenamefont
  {Vallone}, \citenamefont {Villoresi},\ and\ \citenamefont
  {Wallden}}]{Pirandola:AQCrypt2020}%
  \BibitemOpen
  \bibfield  {author} {\bibinfo {author} {\bibfnamefont {S.}~\bibnamefont
  {Pirandola}}, \bibinfo {author} {\bibfnamefont {U.~L.}\ \bibnamefont
  {Andersen}}, \bibinfo {author} {\bibfnamefont {L.}~\bibnamefont {Banchi}},
  \bibinfo {author} {\bibfnamefont {M.}~\bibnamefont {Berta}}, \bibinfo
  {author} {\bibfnamefont {D.}~\bibnamefont {Bunandar}}, \bibinfo {author}
  {\bibfnamefont {R.}~\bibnamefont {Colbeck}}, \bibinfo {author} {\bibfnamefont
  {D.}~\bibnamefont {Englund}}, \bibinfo {author} {\bibfnamefont
  {T.}~\bibnamefont {Gehring}}, \bibinfo {author} {\bibfnamefont
  {C.}~\bibnamefont {Lupo}}, \bibinfo {author} {\bibfnamefont {C.}~\bibnamefont
  {Ottaviani}}, \bibinfo {author} {\bibfnamefont {J.~L.}\ \bibnamefont
  {Pereira}}, \bibinfo {author} {\bibfnamefont {M.}~\bibnamefont {Razavi}},
  \bibinfo {author} {\bibfnamefont {J.~S.}\ \bibnamefont {Shaari}}, \bibinfo
  {author} {\bibfnamefont {M.}~\bibnamefont {Tomamichel}}, \bibinfo {author}
  {\bibfnamefont {V.~C.}\ \bibnamefont {Usenko}}, \bibinfo {author}
  {\bibfnamefont {G.}~\bibnamefont {Vallone}}, \bibinfo {author} {\bibfnamefont
  {P.}~\bibnamefont {Villoresi}}, \ and\ \bibinfo {author} {\bibfnamefont
  {P.}~\bibnamefont {Wallden}},\ }\href {\doibase 10.1364/AOP.361502}
  {\bibfield  {journal} {\bibinfo  {journal} {Adv. Opt. Photon.}\ }\textbf
  {\bibinfo {volume} {12}},\ \bibinfo {pages} {1012} (\bibinfo {year}
  {2020})}\BibitemShut {NoStop}%
\bibitem [{\citenamefont {Colbeck}\ and\ \citenamefont
  {Renner}(2011)}]{Colbeck:QRNGs2012}%
  \BibitemOpen
  \bibfield  {author} {\bibinfo {author} {\bibfnamefont {R.}~\bibnamefont
  {Colbeck}}\ and\ \bibinfo {author} {\bibfnamefont {R.}~\bibnamefont
  {Renner}},\ }\href {\doibase https://doi.org/10.1038/nphys2300} {\bibfield
  {journal} {\bibinfo  {journal} {Nature Physics}\ }\textbf {\bibinfo {volume}
  {8}},\ \bibinfo {pages} {450–453} (\bibinfo {year} {2011})}\BibitemShut
  {NoStop}%
\bibitem [{\citenamefont {Gallego}\ \emph {et~al.}(2013)\citenamefont
  {Gallego}, \citenamefont {Masanes}, \citenamefont {Torre}, \citenamefont
  {Dhara}, \citenamefont {Aolita},\ and\ \citenamefont
  {Acín}}]{Gallego:QRNGs2013}%
  \BibitemOpen
  \bibfield  {author} {\bibinfo {author} {\bibfnamefont {R.}~\bibnamefont
  {Gallego}}, \bibinfo {author} {\bibfnamefont {L.}~\bibnamefont {Masanes}},
  \bibinfo {author} {\bibfnamefont {G.~D.~L.}\ \bibnamefont {Torre}}, \bibinfo
  {author} {\bibfnamefont {C.}~\bibnamefont {Dhara}}, \bibinfo {author}
  {\bibfnamefont {L.}~\bibnamefont {Aolita}}, \ and\ \bibinfo {author}
  {\bibfnamefont {A.}~\bibnamefont {Acín}},\ }\href {\doibase
  https://doi.org/10.1038/ncomms3654} {\bibfield  {journal} {\bibinfo
  {journal} {Nature Communications}\ }\textbf {\bibinfo {volume} {4}},\
  \bibinfo {pages} {2654} (\bibinfo {year} {2013})}\BibitemShut {NoStop}%
\bibitem [{\citenamefont {Mitchell}\ \emph {et~al.}(2015)\citenamefont
  {Mitchell}, \citenamefont {Abellan},\ and\ \citenamefont
  {Amaya}}]{Mitchell:QRNGs2015}%
  \BibitemOpen
  \bibfield  {author} {\bibinfo {author} {\bibfnamefont {M.~W.}\ \bibnamefont
  {Mitchell}}, \bibinfo {author} {\bibfnamefont {C.}~\bibnamefont {Abellan}}, \
  and\ \bibinfo {author} {\bibfnamefont {W.}~\bibnamefont {Amaya}},\ }\href
  {\doibase 10.1103/PhysRevA.91.012314} {\bibfield  {journal} {\bibinfo
  {journal} {Phys. Rev. A}\ }\textbf {\bibinfo {volume} {91}},\ \bibinfo
  {pages} {012314} (\bibinfo {year} {2015})}\BibitemShut {NoStop}%
\bibitem [{\citenamefont {Lunghi}\ \emph {et~al.}(2015)\citenamefont {Lunghi},
  \citenamefont {Brask}, \citenamefont {Lim}, \citenamefont {Lavigne},
  \citenamefont {Bowles}, \citenamefont {Martin}, \citenamefont {Zbinden},\
  and\ \citenamefont {Brunner}}]{Lunghi:QRNGs2015}%
  \BibitemOpen
  \bibfield  {author} {\bibinfo {author} {\bibfnamefont {T.}~\bibnamefont
  {Lunghi}}, \bibinfo {author} {\bibfnamefont {J.~B.}\ \bibnamefont {Brask}},
  \bibinfo {author} {\bibfnamefont {C.~C.~W.}\ \bibnamefont {Lim}}, \bibinfo
  {author} {\bibfnamefont {Q.}~\bibnamefont {Lavigne}}, \bibinfo {author}
  {\bibfnamefont {J.}~\bibnamefont {Bowles}}, \bibinfo {author} {\bibfnamefont
  {A.}~\bibnamefont {Martin}}, \bibinfo {author} {\bibfnamefont
  {H.}~\bibnamefont {Zbinden}}, \ and\ \bibinfo {author} {\bibfnamefont
  {N.}~\bibnamefont {Brunner}},\ }\href {\doibase
  10.1103/PhysRevLett.114.150501} {\bibfield  {journal} {\bibinfo  {journal}
  {Phys. Rev. Lett.}\ }\textbf {\bibinfo {volume} {114}},\ \bibinfo {pages}
  {150501} (\bibinfo {year} {2015})}\BibitemShut {NoStop}%
\bibitem [{\citenamefont {Brandão}\ \emph {et~al.}(2016)\citenamefont
  {Brandão}, \citenamefont {Ramanathan}, \citenamefont {Grudka}, \citenamefont
  {Horodecki}, \citenamefont {Horodecki}, \citenamefont {Horodecki},
  \citenamefont {Szarek},\ and\ \citenamefont
  {Wojewódka}}]{Brandao:QRNGs2016}%
  \BibitemOpen
  \bibfield  {author} {\bibinfo {author} {\bibfnamefont {F.~G. S.~L.}\
  \bibnamefont {Brandão}}, \bibinfo {author} {\bibfnamefont {R.}~\bibnamefont
  {Ramanathan}}, \bibinfo {author} {\bibfnamefont {A.}~\bibnamefont {Grudka}},
  \bibinfo {author} {\bibfnamefont {K.}~\bibnamefont {Horodecki}}, \bibinfo
  {author} {\bibfnamefont {M.}~\bibnamefont {Horodecki}}, \bibinfo {author}
  {\bibfnamefont {P.}~\bibnamefont {Horodecki}}, \bibinfo {author}
  {\bibfnamefont {T.}~\bibnamefont {Szarek}}, \ and\ \bibinfo {author}
  {\bibfnamefont {H.}~\bibnamefont {Wojewódka}},\ }\href {\doibase
  https://doi.org/10.1038/ncomms11345} {\bibfield  {journal} {\bibinfo
  {journal} {Nature Communications}\ }\textbf {\bibinfo {volume} {7}},\
  \bibinfo {pages} {11345} (\bibinfo {year} {2016})}\BibitemShut {NoStop}%
\bibitem [{\citenamefont {Ma}\ \emph {et~al.}(2016)\citenamefont {Ma},
  \citenamefont {Yuan}, \citenamefont {Cao}, \citenamefont {Qi},\ and\
  \citenamefont {Zhang}}]{Ma:QRNGs2016}%
  \BibitemOpen
  \bibfield  {author} {\bibinfo {author} {\bibfnamefont {X.}~\bibnamefont
  {Ma}}, \bibinfo {author} {\bibfnamefont {X.}~\bibnamefont {Yuan}}, \bibinfo
  {author} {\bibfnamefont {Z.}~\bibnamefont {Cao}}, \bibinfo {author}
  {\bibfnamefont {B.}~\bibnamefont {Qi}}, \ and\ \bibinfo {author}
  {\bibfnamefont {Z.}~\bibnamefont {Zhang}},\ }\href {\doibase
  https://doi.org/10.1038/npjqi.2016.21} {\bibfield  {journal} {\bibinfo
  {journal} {npj Quantum Information}\ }\textbf {\bibinfo {volume} {2}},\
  \bibinfo {pages} {16021} (\bibinfo {year} {2016})}\BibitemShut {NoStop}%
\bibitem [{\citenamefont {Acín}\ and\ \citenamefont
  {Masanes}(2016)}]{Acin:QRNGs2016}%
  \BibitemOpen
  \bibfield  {author} {\bibinfo {author} {\bibfnamefont {A.}~\bibnamefont
  {Acín}}\ and\ \bibinfo {author} {\bibfnamefont {L.}~\bibnamefont
  {Masanes}},\ }\href {\doibase https://doi.org/10.1038/nature20119} {\bibfield
   {journal} {\bibinfo  {journal} {Nature}\ }\textbf {\bibinfo {volume}
  {540}},\ \bibinfo {pages} {213–219} (\bibinfo {year} {2016})}\BibitemShut
  {NoStop}%
\bibitem [{\citenamefont {Herrero-Collantes}\ and\ \citenamefont
  {Garcia-Escartin}(2017)}]{Herrero-Collantes:QRNGs2017}%
  \BibitemOpen
  \bibfield  {author} {\bibinfo {author} {\bibfnamefont {M.}~\bibnamefont
  {Herrero-Collantes}}\ and\ \bibinfo {author} {\bibfnamefont {J.~C.}\
  \bibnamefont {Garcia-Escartin}},\ }\href {\doibase
  10.1103/RevModPhys.89.015004} {\bibfield  {journal} {\bibinfo  {journal}
  {Rev. Mod. Phys.}\ }\textbf {\bibinfo {volume} {89}},\ \bibinfo {pages}
  {015004} (\bibinfo {year} {2017})}\BibitemShut {NoStop}%
\bibitem [{\citenamefont {Hillery}\ \emph {et~al.}(1999)\citenamefont
  {Hillery}, \citenamefont {Bu\ifmmode~\check{z}\else \v{z}\fi{}ek},\ and\
  \citenamefont {Berthiaume}}]{Hillery:QSS1999}%
  \BibitemOpen
  \bibfield  {author} {\bibinfo {author} {\bibfnamefont {M.}~\bibnamefont
  {Hillery}}, \bibinfo {author} {\bibfnamefont {V.}~\bibnamefont
  {Bu\ifmmode~\check{z}\else \v{z}\fi{}ek}}, \ and\ \bibinfo {author}
  {\bibfnamefont {A.}~\bibnamefont {Berthiaume}},\ }\href {\doibase
  10.1103/PhysRevA.59.1829} {\bibfield  {journal} {\bibinfo  {journal} {Phys.
  Rev. A}\ }\textbf {\bibinfo {volume} {59}},\ \bibinfo {pages} {1829}
  (\bibinfo {year} {1999})}\BibitemShut {NoStop}%
\bibitem [{\citenamefont {Cleve}\ \emph {et~al.}(1999)\citenamefont {Cleve},
  \citenamefont {Gottesman},\ and\ \citenamefont {Lo}}]{Cleve:QSS1999}%
  \BibitemOpen
  \bibfield  {author} {\bibinfo {author} {\bibfnamefont {R.}~\bibnamefont
  {Cleve}}, \bibinfo {author} {\bibfnamefont {D.}~\bibnamefont {Gottesman}}, \
  and\ \bibinfo {author} {\bibfnamefont {H.-K.}\ \bibnamefont {Lo}},\ }\href
  {\doibase 10.1103/PhysRevLett.83.648} {\bibfield  {journal} {\bibinfo
  {journal} {Phys. Rev. Lett.}\ }\textbf {\bibinfo {volume} {83}},\ \bibinfo
  {pages} {648} (\bibinfo {year} {1999})}\BibitemShut {NoStop}%
\bibitem [{\citenamefont {Ribeiro}\ \emph {et~al.}(2018)\citenamefont
  {Ribeiro}, \citenamefont {Murta},\ and\ \citenamefont
  {Wehner}}]{Ribeiro:QSS2010}%
  \BibitemOpen
  \bibfield  {author} {\bibinfo {author} {\bibfnamefont {J.}~\bibnamefont
  {Ribeiro}}, \bibinfo {author} {\bibfnamefont {G.}~\bibnamefont {Murta}}, \
  and\ \bibinfo {author} {\bibfnamefont {S.}~\bibnamefont {Wehner}},\ }\href
  {\doibase 10.1103/PhysRevA.97.022307} {\bibfield  {journal} {\bibinfo
  {journal} {Phys. Rev. A}\ }\textbf {\bibinfo {volume} {97}},\ \bibinfo
  {pages} {022307} (\bibinfo {year} {2018})}\BibitemShut {NoStop}%
\bibitem [{\citenamefont {Keet}\ \emph {et~al.}(2010)\citenamefont {Keet},
  \citenamefont {Fortescue}, \citenamefont {Markham},\ and\ \citenamefont
  {Sanders}}]{Keet:QSS2010}%
  \BibitemOpen
  \bibfield  {author} {\bibinfo {author} {\bibfnamefont {A.}~\bibnamefont
  {Keet}}, \bibinfo {author} {\bibfnamefont {B.}~\bibnamefont {Fortescue}},
  \bibinfo {author} {\bibfnamefont {D.}~\bibnamefont {Markham}}, \ and\
  \bibinfo {author} {\bibfnamefont {B.~C.}\ \bibnamefont {Sanders}},\ }\href
  {\doibase 10.1103/PhysRevA.82.062315} {\bibfield  {journal} {\bibinfo
  {journal} {Phys. Rev. A}\ }\textbf {\bibinfo {volume} {82}},\ \bibinfo
  {pages} {062315} (\bibinfo {year} {2010})}\BibitemShut {NoStop}%
\bibitem [{\citenamefont {Ottaviani}\ \emph {et~al.}(2019)\citenamefont
  {Ottaviani}, \citenamefont {Lupo}, \citenamefont {Laurenza},\ and\
  \citenamefont {Pirandola}}]{Ottaviani:QConf2019}%
  \BibitemOpen
  \bibfield  {author} {\bibinfo {author} {\bibfnamefont {C.}~\bibnamefont
  {Ottaviani}}, \bibinfo {author} {\bibfnamefont {C.}~\bibnamefont {Lupo}},
  \bibinfo {author} {\bibfnamefont {R.}~\bibnamefont {Laurenza}}, \ and\
  \bibinfo {author} {\bibfnamefont {S.}~\bibnamefont {Pirandola}},\ }\href
  {\doibase https://doi.org/10.1038/s42005-019-0209-6} {\bibfield  {journal}
  {\bibinfo  {journal} {Communications Physics}\ }\textbf {\bibinfo {volume}
  {2}},\ \bibinfo {pages} {118} (\bibinfo {year} {2019})}\BibitemShut {NoStop}%
\bibitem [{\citenamefont {Grasselli}\ \emph {et~al.}(2019)\citenamefont
  {Grasselli}, \citenamefont {Kampermann},\ and\ \citenamefont
  {Bru{\ss}}}]{Grasselli:QSS2019}%
  \BibitemOpen
  \bibfield  {author} {\bibinfo {author} {\bibfnamefont {F.}~\bibnamefont
  {Grasselli}}, \bibinfo {author} {\bibfnamefont {H.}~\bibnamefont
  {Kampermann}}, \ and\ \bibinfo {author} {\bibfnamefont {D.}~\bibnamefont
  {Bru{\ss}}},\ }\href {\doibase 10.1088/1367-2630/ab573e} {\bibfield
  {journal} {\bibinfo  {journal} {New Journal of Physics}\ }\textbf {\bibinfo
  {volume} {21}},\ \bibinfo {pages} {123002} (\bibinfo {year}
  {2019})}\BibitemShut {NoStop}%
\bibitem [{\citenamefont {Pirandola}(2020)}]{Pirandola:QConf2020}%
  \BibitemOpen
  \bibfield  {author} {\bibinfo {author} {\bibfnamefont {S.}~\bibnamefont
  {Pirandola}},\ }\href {\doibase https://doi.org/10.1049/iet-qtc.2020.0006}
  {\bibfield  {journal} {\bibinfo  {journal} {IET Quantum Communication}\
  }\textbf {\bibinfo {volume} {1}},\ \bibinfo {pages} {22} (\bibinfo {year}
  {2020})}\BibitemShut {NoStop}%
\bibitem [{\citenamefont {Gottesman}\ and\ \citenamefont
  {Chuang}(2001)}]{Gottesman:QDS2001}%
  \BibitemOpen
  \bibfield  {author} {\bibinfo {author} {\bibfnamefont {D.}~\bibnamefont
  {Gottesman}}\ and\ \bibinfo {author} {\bibfnamefont {I.}~\bibnamefont
  {Chuang}},\ }\href@noop {} {\bibfield  {journal} {\bibinfo  {journal}
  {quant-ph/0105032}\ } (\bibinfo {year} {2001})}\BibitemShut {NoStop}%
\bibitem [{\citenamefont {Clarke}\ \emph {et~al.}(2012)\citenamefont {Clarke},
  \citenamefont {Collins}, \citenamefont {andErika Andersson}, \citenamefont
  {Jeffers},\ and\ \citenamefont {Buller}}]{Clarke:QDS2012}%
  \BibitemOpen
  \bibfield  {author} {\bibinfo {author} {\bibfnamefont {P.~J.}\ \bibnamefont
  {Clarke}}, \bibinfo {author} {\bibfnamefont {R.~J.}\ \bibnamefont {Collins}},
  \bibinfo {author} {\bibfnamefont {V.~D.}\ \bibnamefont {andErika Andersson}},
  \bibinfo {author} {\bibfnamefont {J.}~\bibnamefont {Jeffers}}, \ and\
  \bibinfo {author} {\bibfnamefont {G.~S.}\ \bibnamefont {Buller}},\ }\href
  {\doibase https://doi.org/10.1038/ncomms2172} {\bibfield  {journal} {\bibinfo
   {journal} {Nature Communications}\ }\textbf {\bibinfo {volume} {3}},\
  \bibinfo {pages} {1174} (\bibinfo {year} {2012})}\BibitemShut {NoStop}%
\bibitem [{\citenamefont {Collins}\ \emph {et~al.}(2014)\citenamefont
  {Collins}, \citenamefont {Donaldson}, \citenamefont {Dunjko}, \citenamefont
  {Wallden}, \citenamefont {Clarke}, \citenamefont {Andersson}, \citenamefont
  {Jeffers},\ and\ \citenamefont {Buller}}]{Collins:QDS2014}%
  \BibitemOpen
  \bibfield  {author} {\bibinfo {author} {\bibfnamefont {R.~J.}\ \bibnamefont
  {Collins}}, \bibinfo {author} {\bibfnamefont {R.~J.}\ \bibnamefont
  {Donaldson}}, \bibinfo {author} {\bibfnamefont {V.}~\bibnamefont {Dunjko}},
  \bibinfo {author} {\bibfnamefont {P.}~\bibnamefont {Wallden}}, \bibinfo
  {author} {\bibfnamefont {P.~J.}\ \bibnamefont {Clarke}}, \bibinfo {author}
  {\bibfnamefont {E.}~\bibnamefont {Andersson}}, \bibinfo {author}
  {\bibfnamefont {J.}~\bibnamefont {Jeffers}}, \ and\ \bibinfo {author}
  {\bibfnamefont {G.~S.}\ \bibnamefont {Buller}},\ }\href {\doibase
  10.1103/PhysRevLett.113.040502} {\bibfield  {journal} {\bibinfo  {journal}
  {Phys. Rev. Lett.}\ }\textbf {\bibinfo {volume} {113}},\ \bibinfo {pages}
  {040502} (\bibinfo {year} {2014})}\BibitemShut {NoStop}%
\bibitem [{\citenamefont {Dunjko}\ \emph {et~al.}(2014)\citenamefont {Dunjko},
  \citenamefont {Wallden},\ and\ \citenamefont {Andersson}}]{Dunjko:QDS2014}%
  \BibitemOpen
  \bibfield  {author} {\bibinfo {author} {\bibfnamefont {V.}~\bibnamefont
  {Dunjko}}, \bibinfo {author} {\bibfnamefont {P.}~\bibnamefont {Wallden}}, \
  and\ \bibinfo {author} {\bibfnamefont {E.}~\bibnamefont {Andersson}},\ }\href
  {\doibase 10.1103/PhysRevLett.112.040502} {\bibfield  {journal} {\bibinfo
  {journal} {Phys. Rev. Lett.}\ }\textbf {\bibinfo {volume} {112}},\ \bibinfo
  {pages} {040502} (\bibinfo {year} {2014})}\BibitemShut {NoStop}%
\bibitem [{\citenamefont {Wallden}\ \emph {et~al.}(2015)\citenamefont
  {Wallden}, \citenamefont {Dunjko}, \citenamefont {Kent},\ and\ \citenamefont
  {Andersson}}]{Wallden:QDS2015}%
  \BibitemOpen
  \bibfield  {author} {\bibinfo {author} {\bibfnamefont {P.}~\bibnamefont
  {Wallden}}, \bibinfo {author} {\bibfnamefont {V.}~\bibnamefont {Dunjko}},
  \bibinfo {author} {\bibfnamefont {A.}~\bibnamefont {Kent}}, \ and\ \bibinfo
  {author} {\bibfnamefont {E.}~\bibnamefont {Andersson}},\ }\href {\doibase
  10.1103/PhysRevA.91.042304} {\bibfield  {journal} {\bibinfo  {journal} {Phys.
  Rev. A}\ }\textbf {\bibinfo {volume} {91}},\ \bibinfo {pages} {042304}
  (\bibinfo {year} {2015})}\BibitemShut {NoStop}%
\bibitem [{\citenamefont {Croal}\ \emph {et~al.}(2016)\citenamefont {Croal},
  \citenamefont {Peuntinger}, \citenamefont {Heim}, \citenamefont {Khan},
  \citenamefont {Marquardt}, \citenamefont {Leuchs}, \citenamefont {Wallden},
  \citenamefont {Andersson},\ and\ \citenamefont {Korolkova}}]{Croal:QDS2016}%
  \BibitemOpen
  \bibfield  {author} {\bibinfo {author} {\bibfnamefont {C.}~\bibnamefont
  {Croal}}, \bibinfo {author} {\bibfnamefont {C.}~\bibnamefont {Peuntinger}},
  \bibinfo {author} {\bibfnamefont {B.}~\bibnamefont {Heim}}, \bibinfo {author}
  {\bibfnamefont {I.}~\bibnamefont {Khan}}, \bibinfo {author} {\bibfnamefont
  {C.}~\bibnamefont {Marquardt}}, \bibinfo {author} {\bibfnamefont
  {G.}~\bibnamefont {Leuchs}}, \bibinfo {author} {\bibfnamefont
  {P.}~\bibnamefont {Wallden}}, \bibinfo {author} {\bibfnamefont
  {E.}~\bibnamefont {Andersson}}, \ and\ \bibinfo {author} {\bibfnamefont
  {N.}~\bibnamefont {Korolkova}},\ }\href {\doibase
  10.1103/PhysRevLett.117.100503} {\bibfield  {journal} {\bibinfo  {journal}
  {Phys. Rev. Lett.}\ }\textbf {\bibinfo {volume} {117}},\ \bibinfo {pages}
  {100503} (\bibinfo {year} {2016})}\BibitemShut {NoStop}%
\bibitem [{\citenamefont {Collins}\ \emph {et~al.}(2016)\citenamefont
  {Collins}, \citenamefont {Amiri}, \citenamefont {Fujiwara}, \citenamefont
  {Honjo}, \citenamefont {Shimizu}, \citenamefont {Tamaki}, \citenamefont
  {Takeoka}, \citenamefont {Andersson}, \citenamefont {Buller},\ and\
  \citenamefont {Sasaki}}]{Collins:QDS2016}%
  \BibitemOpen
  \bibfield  {author} {\bibinfo {author} {\bibfnamefont {R.~J.}\ \bibnamefont
  {Collins}}, \bibinfo {author} {\bibfnamefont {R.}~\bibnamefont {Amiri}},
  \bibinfo {author} {\bibfnamefont {M.}~\bibnamefont {Fujiwara}}, \bibinfo
  {author} {\bibfnamefont {T.}~\bibnamefont {Honjo}}, \bibinfo {author}
  {\bibfnamefont {K.}~\bibnamefont {Shimizu}}, \bibinfo {author} {\bibfnamefont
  {K.}~\bibnamefont {Tamaki}}, \bibinfo {author} {\bibfnamefont
  {M.}~\bibnamefont {Takeoka}}, \bibinfo {author} {\bibfnamefont
  {E.}~\bibnamefont {Andersson}}, \bibinfo {author} {\bibfnamefont {G.~S.}\
  \bibnamefont {Buller}}, \ and\ \bibinfo {author} {\bibfnamefont
  {M.}~\bibnamefont {Sasaki}},\ }\href {\doibase 10.1364/OL.41.004883}
  {\bibfield  {journal} {\bibinfo  {journal} {Opt. Lett.}\ }\textbf {\bibinfo
  {volume} {41}},\ \bibinfo {pages} {4883} (\bibinfo {year}
  {2016})}\BibitemShut {NoStop}%
\bibitem [{\citenamefont {Roberts}\ \emph {et~al.}(2017)\citenamefont
  {Roberts}, \citenamefont {Lucamarini}, \citenamefont {Yuan}, \citenamefont
  {Dynes}, \citenamefont {Comandar}, \citenamefont {Sharpe}, \citenamefont
  {Shields}, \citenamefont {Curty}, \citenamefont {Puthoor},\ and\
  \citenamefont {Andersson}}]{Roberts:QDS2017}%
  \BibitemOpen
  \bibfield  {author} {\bibinfo {author} {\bibfnamefont {G.~L.}\ \bibnamefont
  {Roberts}}, \bibinfo {author} {\bibfnamefont {M.}~\bibnamefont {Lucamarini}},
  \bibinfo {author} {\bibfnamefont {Z.~L.}\ \bibnamefont {Yuan}}, \bibinfo
  {author} {\bibfnamefont {J.~F.}\ \bibnamefont {Dynes}}, \bibinfo {author}
  {\bibfnamefont {L.~C.}\ \bibnamefont {Comandar}}, \bibinfo {author}
  {\bibfnamefont {A.~W.}\ \bibnamefont {Sharpe}}, \bibinfo {author}
  {\bibfnamefont {A.~J.}\ \bibnamefont {Shields}}, \bibinfo {author}
  {\bibfnamefont {M.}~\bibnamefont {Curty}}, \bibinfo {author} {\bibfnamefont
  {I.~V.}\ \bibnamefont {Puthoor}}, \ and\ \bibinfo {author} {\bibfnamefont
  {E.}~\bibnamefont {Andersson}},\ }\href {\doibase
  https://doi.org/10.1038/s41467-017-01245-5} {\bibfield  {journal} {\bibinfo
  {journal} {Nature Communications}\ }\textbf {\bibinfo {volume} {8}},\
  \bibinfo {pages} {1098} (\bibinfo {year} {2017})}\BibitemShut {NoStop}%
\bibitem [{\citenamefont {Mayers}(1997)}]{Mayers:BitCom1997}%
  \BibitemOpen
  \bibfield  {author} {\bibinfo {author} {\bibfnamefont {D.}~\bibnamefont
  {Mayers}},\ }\href {\doibase 10.1103/PhysRevLett.78.3414} {\bibfield
  {journal} {\bibinfo  {journal} {Phys. Rev. Lett.}\ }\textbf {\bibinfo
  {volume} {78}},\ \bibinfo {pages} {3414} (\bibinfo {year}
  {1997})}\BibitemShut {NoStop}%
\bibitem [{\citenamefont {Lo}\ and\ \citenamefont
  {Chau}(1998)}]{LO:BitCom1998}%
  \BibitemOpen
  \bibfield  {author} {\bibinfo {author} {\bibfnamefont {H.-K.}\ \bibnamefont
  {Lo}}\ and\ \bibinfo {author} {\bibfnamefont {H.}~\bibnamefont {Chau}},\
  }\href {\doibase https://doi.org/10.1016/S0167-2789(98)00053-0} {\bibfield
  {journal} {\bibinfo  {journal} {Physica D: Nonlinear Phenomena}\ }\textbf
  {\bibinfo {volume} {120}},\ \bibinfo {pages} {177} (\bibinfo {year}
  {1998})},\ \bibinfo {note} {proceedings of the Fourth Workshop on Physics and
  Consumption}\BibitemShut {NoStop}%
\bibitem [{\citenamefont {Buhrman}\ \emph {et~al.}(2008)\citenamefont
  {Buhrman}, \citenamefont {Christandl}, \citenamefont {Hayden}, \citenamefont
  {Lo},\ and\ \citenamefont {Wehner}}]{Buhrman::BitCom2008}%
  \BibitemOpen
  \bibfield  {author} {\bibinfo {author} {\bibfnamefont {H.}~\bibnamefont
  {Buhrman}}, \bibinfo {author} {\bibfnamefont {M.}~\bibnamefont {Christandl}},
  \bibinfo {author} {\bibfnamefont {P.}~\bibnamefont {Hayden}}, \bibinfo
  {author} {\bibfnamefont {H.-K.}\ \bibnamefont {Lo}}, \ and\ \bibinfo {author}
  {\bibfnamefont {S.}~\bibnamefont {Wehner}},\ }\href {\doibase
  10.1103/PhysRevA.78.022316} {\bibfield  {journal} {\bibinfo  {journal} {Phys.
  Rev. A}\ }\textbf {\bibinfo {volume} {78}},\ \bibinfo {pages} {022316}
  (\bibinfo {year} {2008})}\BibitemShut {NoStop}%
\bibitem [{\citenamefont {Chailloux}\ and\ \citenamefont
  {Kerenidis}(2011)}]{Chailloux:BitCom2011}%
  \BibitemOpen
  \bibfield  {author} {\bibinfo {author} {\bibfnamefont {A.}~\bibnamefont
  {Chailloux}}\ and\ \bibinfo {author} {\bibfnamefont {I.}~\bibnamefont
  {Kerenidis}},\ }in\ \href {\doibase 10.1109/FOCS.2011.42} {\emph {\bibinfo
  {booktitle} {2011 IEEE 52nd Annual Symposium on Foundations of Computer
  Science}}}\ (\bibinfo {year} {2011})\ pp.\ \bibinfo {pages}
  {354--362}\BibitemShut {NoStop}%
\bibitem [{\citenamefont {Kent}(2012)}]{Kent:BitCom2012}%
  \BibitemOpen
  \bibfield  {author} {\bibinfo {author} {\bibfnamefont {A.}~\bibnamefont
  {Kent}},\ }\href {\doibase 10.1103/PhysRevLett.109.130501} {\bibfield
  {journal} {\bibinfo  {journal} {Phys. Rev. Lett.}\ }\textbf {\bibinfo
  {volume} {109}},\ \bibinfo {pages} {130501} (\bibinfo {year}
  {2012})}\BibitemShut {NoStop}%
\bibitem [{\citenamefont {Broadbent}\ \emph {et~al.}(2009)\citenamefont
  {Broadbent}, \citenamefont {Fitzsimons},\ and\ \citenamefont
  {Kashefi}}]{Broadbent:SQC2009}%
  \BibitemOpen
  \bibfield  {author} {\bibinfo {author} {\bibfnamefont {A.}~\bibnamefont
  {Broadbent}}, \bibinfo {author} {\bibfnamefont {J.}~\bibnamefont
  {Fitzsimons}}, \ and\ \bibinfo {author} {\bibfnamefont {E.}~\bibnamefont
  {Kashefi}},\ }in\ \href {\doibase 10.1109/FOCS.2009.36} {\emph {\bibinfo
  {booktitle} {2009 50th Annual IEEE Symposium on Foundations of Computer
  Science}}}\ (\bibinfo {year} {2009})\ pp.\ \bibinfo {pages}
  {517--526}\BibitemShut {NoStop}%
\bibitem [{\citenamefont {Barz}\ \emph {et~al.}(2012)\citenamefont {Barz},
  \citenamefont {Kashefi}, \citenamefont {Broadbent}, \citenamefont
  {Fitzsimons}, \citenamefont {Zeilinger},\ and\ \citenamefont
  {Walther}}]{Barz:SQC2012}%
  \BibitemOpen
  \bibfield  {author} {\bibinfo {author} {\bibfnamefont {S.}~\bibnamefont
  {Barz}}, \bibinfo {author} {\bibfnamefont {E.}~\bibnamefont {Kashefi}},
  \bibinfo {author} {\bibfnamefont {A.}~\bibnamefont {Broadbent}}, \bibinfo
  {author} {\bibfnamefont {J.~F.}\ \bibnamefont {Fitzsimons}}, \bibinfo
  {author} {\bibfnamefont {A.}~\bibnamefont {Zeilinger}}, \ and\ \bibinfo
  {author} {\bibfnamefont {P.}~\bibnamefont {Walther}},\ }\href {\doibase
  10.1126/science.1214707} {\bibfield  {journal} {\bibinfo  {journal}
  {Science}\ }\textbf {\bibinfo {volume} {335}},\ \bibinfo {pages} {303}
  (\bibinfo {year} {2012})}\BibitemShut {NoStop}%
\bibitem [{\citenamefont {Barz}\ \emph {et~al.}(2013)\citenamefont {Barz},
  \citenamefont {Fitzsimons}, \citenamefont {Kashefi},\ and\ \citenamefont
  {Walther}}]{Barz:SQC2013}%
  \BibitemOpen
  \bibfield  {author} {\bibinfo {author} {\bibfnamefont {S.}~\bibnamefont
  {Barz}}, \bibinfo {author} {\bibfnamefont {J.~F.}\ \bibnamefont
  {Fitzsimons}}, \bibinfo {author} {\bibfnamefont {E.}~\bibnamefont {Kashefi}},
  \ and\ \bibinfo {author} {\bibfnamefont {P.}~\bibnamefont {Walther}},\ }\href
  {\doibase https://doi.org/10.1038/nphys2763} {\bibfield  {journal} {\bibinfo
  {journal} {Nature Physics}\ }\textbf {\bibinfo {volume} {9}},\ \bibinfo
  {pages} {727–731} (\bibinfo {year} {2013})}\BibitemShut {NoStop}%
\bibitem [{\citenamefont {Fitzsimons}\ and\ \citenamefont
  {Kashefi}(2017)}]{Fitzsimons:SQC2017}%
  \BibitemOpen
  \bibfield  {author} {\bibinfo {author} {\bibfnamefont {J.~F.}\ \bibnamefont
  {Fitzsimons}}\ and\ \bibinfo {author} {\bibfnamefont {E.}~\bibnamefont
  {Kashefi}},\ }\href {\doibase 10.1103/PhysRevA.96.012303} {\bibfield
  {journal} {\bibinfo  {journal} {Phys. Rev. A}\ }\textbf {\bibinfo {volume}
  {96}},\ \bibinfo {pages} {012303} (\bibinfo {year} {2017})}\BibitemShut
  {NoStop}%
\bibitem [{\citenamefont {Kashefi}\ and\ \citenamefont
  {Pappa}(2017)}]{Kashefi:SQC2017}%
  \BibitemOpen
  \bibfield  {author} {\bibinfo {author} {\bibfnamefont {E.}~\bibnamefont
  {Kashefi}}\ and\ \bibinfo {author} {\bibfnamefont {A.}~\bibnamefont
  {Pappa}},\ }\href {https://www.mdpi.com/2410-387X/1/2/12} {\bibfield
  {journal} {\bibinfo  {journal} {Cryptography}\ }\textbf {\bibinfo {volume}
  {1}},\ \bibinfo {pages} {2} (\bibinfo {year} {2017})}\BibitemShut {NoStop}%
\bibitem [{\citenamefont {Kimble}(2008)}]{Kimble:QInt2008}%
  \BibitemOpen
  \bibfield  {author} {\bibinfo {author} {\bibfnamefont {H.~J.}\ \bibnamefont
  {Kimble}},\ }\href {\doibase 10.1038/nature07127} {\bibfield  {journal}
  {\bibinfo  {journal} {Nature}\ }\textbf {\bibinfo {volume} {453}},\ \bibinfo
  {pages} {1023} (\bibinfo {year} {2008})}\BibitemShut {NoStop}%
\bibitem [{\citenamefont {Azuma}\ \emph {et~al.}(2016)\citenamefont {Azuma},
  \citenamefont {Mizutani},\ and\ \citenamefont {Lo}}]{Azuma:QNet2016}%
  \BibitemOpen
  \bibfield  {author} {\bibinfo {author} {\bibfnamefont {K.}~\bibnamefont
  {Azuma}}, \bibinfo {author} {\bibfnamefont {A.}~\bibnamefont {Mizutani}}, \
  and\ \bibinfo {author} {\bibfnamefont {H.-K.}\ \bibnamefont {Lo}},\ }\href
  {\doibase 10.1038/532169a} {\bibfield  {journal} {\bibinfo  {journal} {Nature
  Communications}\ }\textbf {\bibinfo {volume} {7}},\ \bibinfo {pages} {13523}
  (\bibinfo {year} {2016})}\BibitemShut {NoStop}%
\bibitem [{\citenamefont {Pirandola}\ and\ \citenamefont
  {Braunstein}(2016)}]{Pirandola:QNet2016}%
  \BibitemOpen
  \bibfield  {author} {\bibinfo {author} {\bibfnamefont {S.}~\bibnamefont
  {Pirandola}}\ and\ \bibinfo {author} {\bibfnamefont {S.~L.}\ \bibnamefont
  {Braunstein}},\ }\href {\doibase 10.1038/532169a} {\bibfield  {journal}
  {\bibinfo  {journal} {Nature}\ }\textbf {\bibinfo {volume} {532}},\ \bibinfo
  {pages} {169} (\bibinfo {year} {2016})}\BibitemShut {NoStop}%
\bibitem [{\citenamefont {Wehner}\ \emph {et~al.}(2018)\citenamefont {Wehner},
  \citenamefont {Elkouss},\ and\ \citenamefont {Hanson}}]{Barz:QNet2018}%
  \BibitemOpen
  \bibfield  {author} {\bibinfo {author} {\bibfnamefont {S.}~\bibnamefont
  {Wehner}}, \bibinfo {author} {\bibfnamefont {D.}~\bibnamefont {Elkouss}}, \
  and\ \bibinfo {author} {\bibfnamefont {R.}~\bibnamefont {Hanson}},\ }\href
  {\doibase 10.1126/science.aam9288} {\bibfield  {journal} {\bibinfo  {journal}
  {Science}\ }\textbf {\bibinfo {volume} {362}},\ \bibinfo {pages} {6412}
  (\bibinfo {year} {2018})}\BibitemShut {NoStop}%
\bibitem [{\citenamefont {Azuma}(2019)}]{Azuma:QInt2019}%
  \BibitemOpen
  \bibfield  {author} {\bibinfo {author} {\bibfnamefont {K.}~\bibnamefont
  {Azuma}},\ }\href@noop {} {\bibfield  {journal} {\bibinfo  {journal} {De
  Physicus,}\ }\bibinfo {series} {pages: 52-54} (\bibinfo {year}
  {2019})}\BibitemShut {NoStop}%
\bibitem [{\citenamefont {Ekert}(1991)}]{Ekert:1991}%
  \BibitemOpen
  \bibfield  {author} {\bibinfo {author} {\bibfnamefont {A.~K.}\ \bibnamefont
  {Ekert}},\ }\href {\doibase 10.1103/PhysRevLett.67.661} {\bibfield  {journal}
  {\bibinfo  {journal} {Phys. Rev. Lett.}\ }\textbf {\bibinfo {volume} {67}},\
  \bibinfo {pages} {661} (\bibinfo {year} {1991})}\BibitemShut {NoStop}%
\bibitem [{\citenamefont {Bennett}(1992)}]{Bennett:1992}%
  \BibitemOpen
  \bibfield  {author} {\bibinfo {author} {\bibfnamefont {C.~H.}\ \bibnamefont
  {Bennett}},\ }\href {\doibase 10.1103/PhysRevLett.68.3121} {\bibfield
  {journal} {\bibinfo  {journal} {Phys. Rev. Lett.}\ }\textbf {\bibinfo
  {volume} {68}},\ \bibinfo {pages} {3121} (\bibinfo {year}
  {1992})}\BibitemShut {NoStop}%
\bibitem [{\citenamefont {Bru\ss{}}(1998)}]{Bruss:SixSt1998}%
  \BibitemOpen
  \bibfield  {author} {\bibinfo {author} {\bibfnamefont {D.}~\bibnamefont
  {Bru\ss{}}},\ }\href {\doibase 10.1103/PhysRevLett.81.3018} {\bibfield
  {journal} {\bibinfo  {journal} {Phys. Rev. Lett.}\ }\textbf {\bibinfo
  {volume} {81}},\ \bibinfo {pages} {3018} (\bibinfo {year}
  {1998})}\BibitemShut {NoStop}%
\bibitem [{\citenamefont {Weedbrook}\ \emph
  {et~al.}(2012{\natexlab{a}})\citenamefont {Weedbrook}, \citenamefont
  {Pirandola}, \citenamefont {Garc\'{\i}a-Patr\'on}, \citenamefont {Cerf},
  \citenamefont {Ralph}, \citenamefont {Shapiro},\ and\ \citenamefont
  {Lloyd}}]{Weedbrook:RMP2012}%
  \BibitemOpen
  \bibfield  {author} {\bibinfo {author} {\bibfnamefont {C.}~\bibnamefont
  {Weedbrook}}, \bibinfo {author} {\bibfnamefont {S.}~\bibnamefont
  {Pirandola}}, \bibinfo {author} {\bibfnamefont {R.}~\bibnamefont
  {Garc\'{\i}a-Patr\'on}}, \bibinfo {author} {\bibfnamefont {N.~J.}\
  \bibnamefont {Cerf}}, \bibinfo {author} {\bibfnamefont {T.~C.}\ \bibnamefont
  {Ralph}}, \bibinfo {author} {\bibfnamefont {J.~H.}\ \bibnamefont {Shapiro}},
  \ and\ \bibinfo {author} {\bibfnamefont {S.}~\bibnamefont {Lloyd}},\ }\href
  {\doibase 10.1103/RevModPhys.84.621} {\bibfield  {journal} {\bibinfo
  {journal} {Rev. Mod. Phys.}\ }\textbf {\bibinfo {volume} {84}},\ \bibinfo
  {pages} {621} (\bibinfo {year} {2012}{\natexlab{a}})}\BibitemShut {NoStop}%
\bibitem [{\citenamefont {Ralph}(1999)}]{Ralph:CVQKD-Sq1999}%
  \BibitemOpen
  \bibfield  {author} {\bibinfo {author} {\bibfnamefont {T.~C.}\ \bibnamefont
  {Ralph}},\ }\href {\doibase 10.1103/PhysRevA.61.010303} {\bibfield  {journal}
  {\bibinfo  {journal} {Phys. Rev. A}\ }\textbf {\bibinfo {volume} {61}},\
  \bibinfo {pages} {010303} (\bibinfo {year} {1999})}\BibitemShut {NoStop}%
\bibitem [{\citenamefont {Hillery}(2000)}]{Hillery:CVQKD-Sq2000}%
  \BibitemOpen
  \bibfield  {author} {\bibinfo {author} {\bibfnamefont {M.}~\bibnamefont
  {Hillery}},\ }\href {\doibase 10.1103/PhysRevA.61.022309} {\bibfield
  {journal} {\bibinfo  {journal} {Phys. Rev. A}\ }\textbf {\bibinfo {volume}
  {61}},\ \bibinfo {pages} {022309} (\bibinfo {year} {2000})}\BibitemShut
  {NoStop}%
\bibitem [{\citenamefont {Reid}(2000)}]{Reid:CVQKD-Sq2000}%
  \BibitemOpen
  \bibfield  {author} {\bibinfo {author} {\bibfnamefont {M.~D.}\ \bibnamefont
  {Reid}},\ }\href {\doibase 10.1103/PhysRevA.62.062308} {\bibfield  {journal}
  {\bibinfo  {journal} {Phys. Rev. A}\ }\textbf {\bibinfo {volume} {62}},\
  \bibinfo {pages} {062308} (\bibinfo {year} {2000})}\BibitemShut {NoStop}%
\bibitem [{\citenamefont {Grosshans}\ and\ \citenamefont
  {Grangier}(2002)}]{Grosshans:PRL2002}%
  \BibitemOpen
  \bibfield  {author} {\bibinfo {author} {\bibfnamefont {F.}~\bibnamefont
  {Grosshans}}\ and\ \bibinfo {author} {\bibfnamefont {P.}~\bibnamefont
  {Grangier}},\ }\href {\doibase 10.1103/PhysRevLett.88.057902} {\bibfield
  {journal} {\bibinfo  {journal} {Phys. Rev. Lett.}\ }\textbf {\bibinfo
  {volume} {88}},\ \bibinfo {pages} {057902} (\bibinfo {year}
  {2002})}\BibitemShut {NoStop}%
\bibitem [{\citenamefont {Grosshans}\ \emph {et~al.}(2003)\citenamefont
  {Grosshans}, \citenamefont {Van~Assche}, \citenamefont {Wenger},
  \citenamefont {Brouri}, \citenamefont {Cerf},\ and\ \citenamefont
  {Grangier}}]{Grosshans:Nat2003}%
  \BibitemOpen
  \bibfield  {author} {\bibinfo {author} {\bibfnamefont {F.}~\bibnamefont
  {Grosshans}}, \bibinfo {author} {\bibfnamefont {G.}~\bibnamefont
  {Van~Assche}}, \bibinfo {author} {\bibfnamefont {J.}~\bibnamefont {Wenger}},
  \bibinfo {author} {\bibfnamefont {R.}~\bibnamefont {Brouri}}, \bibinfo
  {author} {\bibfnamefont {N.~J.}\ \bibnamefont {Cerf}}, \ and\ \bibinfo
  {author} {\bibfnamefont {P.}~\bibnamefont {Grangier}},\ }\href {\doibase
  10.1038/nature01289} {\bibfield  {journal} {\bibinfo  {journal} {Nature}\
  }\textbf {\bibinfo {volume} {421}},\ \bibinfo {pages} {238} (\bibinfo {year}
  {2003})}\BibitemShut {NoStop}%
\bibitem [{\citenamefont {Filip}(2008)}]{Filip:CVQKD-Th2008}%
  \BibitemOpen
  \bibfield  {author} {\bibinfo {author} {\bibfnamefont {R.}~\bibnamefont
  {Filip}},\ }\href {\doibase 10.1103/PhysRevA.77.022310} {\bibfield  {journal}
  {\bibinfo  {journal} {Phys. Rev. A}\ }\textbf {\bibinfo {volume} {77}},\
  \bibinfo {pages} {022310} (\bibinfo {year} {2008})}\BibitemShut {NoStop}%
\bibitem [{\citenamefont {Usenko}\ and\ \citenamefont
  {Filip}(2010)}]{Usenko:CVQKD-Th2010}%
  \BibitemOpen
  \bibfield  {author} {\bibinfo {author} {\bibfnamefont {V.~C.}\ \bibnamefont
  {Usenko}}\ and\ \bibinfo {author} {\bibfnamefont {R.}~\bibnamefont {Filip}},\
  }\href {\doibase 10.1103/PhysRevA.81.022318} {\bibfield  {journal} {\bibinfo
  {journal} {Phys. Rev. A}\ }\textbf {\bibinfo {volume} {81}},\ \bibinfo
  {pages} {022318} (\bibinfo {year} {2010})}\BibitemShut {NoStop}%
\bibitem [{\citenamefont {Weedbrook}\ \emph {et~al.}(2010)\citenamefont
  {Weedbrook}, \citenamefont {Pirandola}, \citenamefont {Lloyd},\ and\
  \citenamefont {Ralph}}]{Weedbrook:CVQKD-Th2010}%
  \BibitemOpen
  \bibfield  {author} {\bibinfo {author} {\bibfnamefont {C.}~\bibnamefont
  {Weedbrook}}, \bibinfo {author} {\bibfnamefont {S.}~\bibnamefont
  {Pirandola}}, \bibinfo {author} {\bibfnamefont {S.}~\bibnamefont {Lloyd}}, \
  and\ \bibinfo {author} {\bibfnamefont {T.~C.}\ \bibnamefont {Ralph}},\ }\href
  {\doibase 10.1103/PhysRevLett.105.110501} {\bibfield  {journal} {\bibinfo
  {journal} {Phys. Rev. Lett.}\ }\textbf {\bibinfo {volume} {105}},\ \bibinfo
  {pages} {110501} (\bibinfo {year} {2010})}\BibitemShut {NoStop}%
\bibitem [{\citenamefont {Weedbrook}\ \emph
  {et~al.}(2012{\natexlab{b}})\citenamefont {Weedbrook}, \citenamefont
  {Pirandola},\ and\ \citenamefont {Ralph}}]{Weedbrook:CVQKD-Th2012}%
  \BibitemOpen
  \bibfield  {author} {\bibinfo {author} {\bibfnamefont {C.}~\bibnamefont
  {Weedbrook}}, \bibinfo {author} {\bibfnamefont {S.}~\bibnamefont
  {Pirandola}}, \ and\ \bibinfo {author} {\bibfnamefont {T.~C.}\ \bibnamefont
  {Ralph}},\ }\href {\doibase 10.1103/PhysRevA.86.022318} {\bibfield  {journal}
  {\bibinfo  {journal} {Phys. Rev. A}\ }\textbf {\bibinfo {volume} {86}},\
  \bibinfo {pages} {022318} (\bibinfo {year} {2012}{\natexlab{b}})}\BibitemShut
  {NoStop}%
\bibitem [{\citenamefont {Weedbrook}\ \emph {et~al.}(2014)\citenamefont
  {Weedbrook}, \citenamefont {Ottaviani},\ and\ \citenamefont
  {Pirandola}}]{Weedbrook:CVQKD-Th2014}%
  \BibitemOpen
  \bibfield  {author} {\bibinfo {author} {\bibfnamefont {C.}~\bibnamefont
  {Weedbrook}}, \bibinfo {author} {\bibfnamefont {C.}~\bibnamefont
  {Ottaviani}}, \ and\ \bibinfo {author} {\bibfnamefont {S.}~\bibnamefont
  {Pirandola}},\ }\href {\doibase 10.1103/PhysRevA.89.012309} {\bibfield
  {journal} {\bibinfo  {journal} {Phys. Rev. A}\ }\textbf {\bibinfo {volume}
  {89}},\ \bibinfo {pages} {012309} (\bibinfo {year} {2014})}\BibitemShut
  {NoStop}%
\bibitem [{\citenamefont {Ottaviani}\ \emph {et~al.}(2020)\citenamefont
  {Ottaviani}, \citenamefont {Woolley}, \citenamefont {Erementchouk},
  \citenamefont {Federici}, \citenamefont {Mazumder}, \citenamefont
  {Pirandola},\ and\ \citenamefont {Weedbrook}}]{Ottaviani:CVQKD-THz2020}%
  \BibitemOpen
  \bibfield  {author} {\bibinfo {author} {\bibfnamefont {C.}~\bibnamefont
  {Ottaviani}}, \bibinfo {author} {\bibfnamefont {M.}~\bibnamefont {Woolley}},
  \bibinfo {author} {\bibfnamefont {M.}~\bibnamefont {Erementchouk}}, \bibinfo
  {author} {\bibfnamefont {J.}~\bibnamefont {Federici}}, \bibinfo {author}
  {\bibfnamefont {P.}~\bibnamefont {Mazumder}}, \bibinfo {author}
  {\bibfnamefont {S.}~\bibnamefont {Pirandola}}, \ and\ \bibinfo {author}
  {\bibfnamefont {C.}~\bibnamefont {Weedbrook}},\ }\href {\doibase
  10.1109/JSAC.2020.2968973} {\bibfield  {journal} {\bibinfo  {journal} {IEEE
  Journal on Selected Areas in Communications}\ }\textbf {\bibinfo {volume}
  {38}},\ \bibinfo {pages} {483} (\bibinfo {year} {2020})}\BibitemShut
  {NoStop}%
\bibitem [{\citenamefont {Weedbrook}\ \emph {et~al.}(2004)\citenamefont
  {Weedbrook}, \citenamefont {Lance}, \citenamefont {Bowen}, \citenamefont
  {Symul}, \citenamefont {Ralph},\ and\ \citenamefont
  {Lam}}]{Weedbrook:CVQKD2004}%
  \BibitemOpen
  \bibfield  {author} {\bibinfo {author} {\bibfnamefont {C.}~\bibnamefont
  {Weedbrook}}, \bibinfo {author} {\bibfnamefont {A.~M.}\ \bibnamefont
  {Lance}}, \bibinfo {author} {\bibfnamefont {W.~P.}\ \bibnamefont {Bowen}},
  \bibinfo {author} {\bibfnamefont {T.}~\bibnamefont {Symul}}, \bibinfo
  {author} {\bibfnamefont {T.~C.}\ \bibnamefont {Ralph}}, \ and\ \bibinfo
  {author} {\bibfnamefont {P.~K.}\ \bibnamefont {Lam}},\ }\href {\doibase
  10.1103/PhysRevLett.93.170504} {\bibfield  {journal} {\bibinfo  {journal}
  {Phys. Rev. Lett.}\ }\textbf {\bibinfo {volume} {93}},\ \bibinfo {pages}
  {170504} (\bibinfo {year} {2004})}\BibitemShut {NoStop}%
\bibitem [{\citenamefont {Leverrier}\ and\ \citenamefont
  {Grangier}(2009)}]{Leverrier:DMCVQKD2009}%
  \BibitemOpen
  \bibfield  {author} {\bibinfo {author} {\bibfnamefont {A.}~\bibnamefont
  {Leverrier}}\ and\ \bibinfo {author} {\bibfnamefont {P.}~\bibnamefont
  {Grangier}},\ }\href {\doibase 10.1103/PhysRevLett.102.180504} {\bibfield
  {journal} {\bibinfo  {journal} {Phys. Rev. Lett.}\ }\textbf {\bibinfo
  {volume} {102}},\ \bibinfo {pages} {180504} (\bibinfo {year}
  {2009})}\BibitemShut {NoStop}%
\bibitem [{\citenamefont {Pirandola}\ \emph {et~al.}(2008)\citenamefont
  {Pirandola}, \citenamefont {Mancini}, \citenamefont {Lloyd},\ and\
  \citenamefont {Braunstein}}]{Pirandola:CV2Way2008}%
  \BibitemOpen
  \bibfield  {author} {\bibinfo {author} {\bibfnamefont {S.}~\bibnamefont
  {Pirandola}}, \bibinfo {author} {\bibfnamefont {S.}~\bibnamefont {Mancini}},
  \bibinfo {author} {\bibfnamefont {S.}~\bibnamefont {Lloyd}}, \ and\ \bibinfo
  {author} {\bibfnamefont {S.~L.}\ \bibnamefont {Braunstein}},\ }\href
  {\doibase https://doi.org/10.1038/nphys1018} {\bibfield  {journal} {\bibinfo
  {journal} {Nature Physics}\ }\textbf {\bibinfo {volume} {4}},\ \bibinfo
  {pages} {726–730} (\bibinfo {year} {2008})}\BibitemShut {NoStop}%
\bibitem [{\citenamefont {Djordjevic}(2020)}]{Djordjevic:hybridQKD20}%
  \BibitemOpen
  \bibfield  {author} {\bibinfo {author} {\bibfnamefont {I.~B.}\ \bibnamefont
  {Djordjevic}},\ }\href {\doibase 10.1109/JPHOT.2019.2946910} {\bibfield
  {journal} {\bibinfo  {journal} {IEEE Photonics Journal}\ }\textbf {\bibinfo
  {volume} {12}},\ \bibinfo {pages} {1} (\bibinfo {year} {2020})}\BibitemShut
  {NoStop}%
\bibitem [{\citenamefont {Djordjevic}(2022)}]{Djordjevic:hybridQKD22}%
  \BibitemOpen
  \bibfield  {author} {\bibinfo {author} {\bibfnamefont {I.~B.}\ \bibnamefont
  {Djordjevic}},\ }\href {\doibase 10.1109/ACCESS.2022.3154468} {\bibfield
  {journal} {\bibinfo  {journal} {IEEE Access}\ }\textbf {\bibinfo {volume}
  {10}},\ \bibinfo {pages} {23284} (\bibinfo {year} {2022})}\BibitemShut
  {NoStop}%
\bibitem [{\citenamefont {Primaatmaja}\ \emph {et~al.}(2022)\citenamefont
  {Primaatmaja}, \citenamefont {Liang}, \citenamefont {Zhang}, \citenamefont
  {Haw}, \citenamefont {Wang},\ and\ \citenamefont
  {Lim}}]{Primaatmaja:hybridQKD}%
  \BibitemOpen
  \bibfield  {author} {\bibinfo {author} {\bibfnamefont {I.~W.}\ \bibnamefont
  {Primaatmaja}}, \bibinfo {author} {\bibfnamefont {C.~C.}\ \bibnamefont
  {Liang}}, \bibinfo {author} {\bibfnamefont {G.}~\bibnamefont {Zhang}},
  \bibinfo {author} {\bibfnamefont {J.~Y.}\ \bibnamefont {Haw}}, \bibinfo
  {author} {\bibfnamefont {C.}~\bibnamefont {Wang}}, \ and\ \bibinfo {author}
  {\bibfnamefont {C.~C.-W.}\ \bibnamefont {Lim}},\ }\href {\doibase
  10.22331/q-2022-01-03-613} {\bibfield  {journal} {\bibinfo  {journal}
  {{Quantum}}\ }\textbf {\bibinfo {volume} {6}},\ \bibinfo {pages} {613}
  (\bibinfo {year} {2022})}\BibitemShut {NoStop}%
\bibitem [{\citenamefont {Mycroft}\ \emph {et~al.}(2023)\citenamefont
  {Mycroft}, \citenamefont {McDermott}, \citenamefont {Buraczewski},\ and\
  \citenamefont {Stobi\ifmmode~\acute{n}\else
  \'{n}\fi{}ska}}]{Mycroft:hybridQKD}%
  \BibitemOpen
  \bibfield  {author} {\bibinfo {author} {\bibfnamefont {M.~E.}\ \bibnamefont
  {Mycroft}}, \bibinfo {author} {\bibfnamefont {T.}~\bibnamefont {McDermott}},
  \bibinfo {author} {\bibfnamefont {A.}~\bibnamefont {Buraczewski}}, \ and\
  \bibinfo {author} {\bibfnamefont {M.}~\bibnamefont
  {Stobi\ifmmode~\acute{n}\else \'{n}\fi{}ska}},\ }\href {\doibase
  10.1103/PhysRevA.107.012607} {\bibfield  {journal} {\bibinfo  {journal}
  {Phys. Rev. A}\ }\textbf {\bibinfo {volume} {107}},\ \bibinfo {pages}
  {012607} (\bibinfo {year} {2023})}\BibitemShut {NoStop}%
\bibitem [{\citenamefont {Pirandola}\ \emph {et~al.}(2017)\citenamefont
  {Pirandola}, \citenamefont {Laurenza}, \citenamefont {Ottaviani},\ and\
  \citenamefont {Banchi}}]{Pirandola:PLOB2017}%
  \BibitemOpen
  \bibfield  {author} {\bibinfo {author} {\bibfnamefont {S.}~\bibnamefont
  {Pirandola}}, \bibinfo {author} {\bibfnamefont {R.}~\bibnamefont {Laurenza}},
  \bibinfo {author} {\bibfnamefont {C.}~\bibnamefont {Ottaviani}}, \ and\
  \bibinfo {author} {\bibfnamefont {L.}~\bibnamefont {Banchi}},\ }\href
  {\doibase https://doi.org/10.1038/ncomms15043} {\bibfield  {journal}
  {\bibinfo  {journal} {Nat. Commun.}\ }\textbf {\bibinfo {volume} {8}},\
  \bibinfo {pages} {15043} (\bibinfo {year} {2017})}\BibitemShut {NoStop}%
\bibitem [{\citenamefont {Pirandola}\ \emph {et~al.}(2009)\citenamefont
  {Pirandola}, \citenamefont {Garc\'{\i}a-Patr\'on}, \citenamefont
  {Braunstein},\ and\ \citenamefont {Lloyd}}]{PirandolaPatron2009}%
  \BibitemOpen
  \bibfield  {author} {\bibinfo {author} {\bibfnamefont {S.}~\bibnamefont
  {Pirandola}}, \bibinfo {author} {\bibfnamefont {R.}~\bibnamefont
  {Garc\'{\i}a-Patr\'on}}, \bibinfo {author} {\bibfnamefont {S.~L.}\
  \bibnamefont {Braunstein}}, \ and\ \bibinfo {author} {\bibfnamefont
  {S.}~\bibnamefont {Lloyd}},\ }\href {\doibase 10.1103/PhysRevLett.102.050503}
  {\bibfield  {journal} {\bibinfo  {journal} {Phys. Rev. Lett.}\ }\textbf
  {\bibinfo {volume} {102}},\ \bibinfo {pages} {050503} (\bibinfo {year}
  {2009})}\BibitemShut {NoStop}%
\bibitem [{\citenamefont
  {Pirandola}(2019{\natexlab{a}})}]{Pirandola:QNet2019a}%
  \BibitemOpen
  \bibfield  {author} {\bibinfo {author} {\bibfnamefont {S.}~\bibnamefont
  {Pirandola}},\ }\href {\doibase 10.1038/s42005-019-0147-3} {\bibfield
  {journal} {\bibinfo  {journal} {Commun. Phys.}\ }\textbf {\bibinfo {volume}
  {2}},\ \bibinfo {pages} {51} (\bibinfo {year} {2019}{\natexlab{a}})},\
  \bibinfo {note} {see also preprint arXiv:1601.00966 (2016).}\BibitemShut
  {Stop}%
\bibitem [{\citenamefont
  {Pirandola}(2019{\natexlab{b}})}]{Pirandola:QNet2019b}%
  \BibitemOpen
  \bibfield  {author} {\bibinfo {author} {\bibfnamefont {S.}~\bibnamefont
  {Pirandola}},\ }\href {\doibase 10.1088/2058-9565/ab3f66} {\bibfield
  {journal} {\bibinfo  {journal} {Quantum Science and Technology}\ }\textbf
  {\bibinfo {volume} {4}},\ \bibinfo {pages} {045006} (\bibinfo {year}
  {2019}{\natexlab{b}})}\BibitemShut {NoStop}%
\bibitem [{\citenamefont {Braunstein}\ and\ \citenamefont
  {Pirandola}(2012)}]{Braunstein:MDI2012}%
  \BibitemOpen
  \bibfield  {author} {\bibinfo {author} {\bibfnamefont {S.~L.}\ \bibnamefont
  {Braunstein}}\ and\ \bibinfo {author} {\bibfnamefont {S.}~\bibnamefont
  {Pirandola}},\ }\href {\doibase 10.1103/PhysRevLett.108.130502} {\bibfield
  {journal} {\bibinfo  {journal} {Phys. Rev. Lett.}\ }\textbf {\bibinfo
  {volume} {108}},\ \bibinfo {pages} {130502} (\bibinfo {year}
  {2012})}\BibitemShut {NoStop}%
\bibitem [{\citenamefont {Lo}\ \emph {et~al.}(2012)\citenamefont {Lo},
  \citenamefont {Curty},\ and\ \citenamefont {Qi}}]{Lo:MDI2012}%
  \BibitemOpen
  \bibfield  {author} {\bibinfo {author} {\bibfnamefont {H.-K.}\ \bibnamefont
  {Lo}}, \bibinfo {author} {\bibfnamefont {M.}~\bibnamefont {Curty}}, \ and\
  \bibinfo {author} {\bibfnamefont {B.}~\bibnamefont {Qi}},\ }\href {\doibase
  10.1103/PhysRevLett.108.130503} {\bibfield  {journal} {\bibinfo  {journal}
  {Phys. Rev. Lett.}\ }\textbf {\bibinfo {volume} {108}},\ \bibinfo {pages}
  {130503} (\bibinfo {year} {2012})}\BibitemShut {NoStop}%
\bibitem [{\citenamefont {Liu}\ \emph {et~al.}(2013)\citenamefont {Liu},
  \citenamefont {Chen}, \citenamefont {Wang}, \citenamefont {Liang},
  \citenamefont {Shentu}, \citenamefont {Wang}, \citenamefont {Cui},
  \citenamefont {Yin}, \citenamefont {Liu}, \citenamefont {Li}, \citenamefont
  {Ma}, \citenamefont {Pelc}, \citenamefont {Fejer}, \citenamefont {Peng},
  \citenamefont {Zhang},\ and\ \citenamefont {Pan}}]{Liu:MDI2012}%
  \BibitemOpen
  \bibfield  {author} {\bibinfo {author} {\bibfnamefont {Y.}~\bibnamefont
  {Liu}}, \bibinfo {author} {\bibfnamefont {T.-Y.}\ \bibnamefont {Chen}},
  \bibinfo {author} {\bibfnamefont {L.-J.}\ \bibnamefont {Wang}}, \bibinfo
  {author} {\bibfnamefont {H.}~\bibnamefont {Liang}}, \bibinfo {author}
  {\bibfnamefont {G.-L.}\ \bibnamefont {Shentu}}, \bibinfo {author}
  {\bibfnamefont {J.}~\bibnamefont {Wang}}, \bibinfo {author} {\bibfnamefont
  {K.}~\bibnamefont {Cui}}, \bibinfo {author} {\bibfnamefont {H.-L.}\
  \bibnamefont {Yin}}, \bibinfo {author} {\bibfnamefont {N.-L.}\ \bibnamefont
  {Liu}}, \bibinfo {author} {\bibfnamefont {L.}~\bibnamefont {Li}}, \bibinfo
  {author} {\bibfnamefont {X.}~\bibnamefont {Ma}}, \bibinfo {author}
  {\bibfnamefont {J.~S.}\ \bibnamefont {Pelc}}, \bibinfo {author}
  {\bibfnamefont {M.~M.}\ \bibnamefont {Fejer}}, \bibinfo {author}
  {\bibfnamefont {C.-Z.}\ \bibnamefont {Peng}}, \bibinfo {author}
  {\bibfnamefont {Q.}~\bibnamefont {Zhang}}, \ and\ \bibinfo {author}
  {\bibfnamefont {J.-W.}\ \bibnamefont {Pan}},\ }\href {\doibase
  10.1103/PhysRevLett.111.130502} {\bibfield  {journal} {\bibinfo  {journal}
  {Phys. Rev. Lett.}\ }\textbf {\bibinfo {volume} {111}},\ \bibinfo {pages}
  {130502} (\bibinfo {year} {2013})}\BibitemShut {NoStop}%
\bibitem [{\citenamefont {Curty}\ \emph {et~al.}(2014)\citenamefont {Curty},
  \citenamefont {Xu}, \citenamefont {Cui}, \citenamefont {Lim}, \citenamefont
  {Tamaki},\ and\ \citenamefont {Lo}}]{Curty:MDI2014}%
  \BibitemOpen
  \bibfield  {author} {\bibinfo {author} {\bibfnamefont {M.}~\bibnamefont
  {Curty}}, \bibinfo {author} {\bibfnamefont {F.}~\bibnamefont {Xu}}, \bibinfo
  {author} {\bibfnamefont {W.}~\bibnamefont {Cui}}, \bibinfo {author}
  {\bibfnamefont {C.~C.~W.}\ \bibnamefont {Lim}}, \bibinfo {author}
  {\bibfnamefont {K.}~\bibnamefont {Tamaki}}, \ and\ \bibinfo {author}
  {\bibfnamefont {H.-K.}\ \bibnamefont {Lo}},\ }\href {\doibase
  https://doi.org/10.1038/ncomms4732} {\bibfield  {journal} {\bibinfo
  {journal} {Nature Communications}\ }\textbf {\bibinfo {volume} {5}},\
  \bibinfo {pages} {3732} (\bibinfo {year} {2014})}\BibitemShut {NoStop}%
\bibitem [{\citenamefont {Lucamarini}\ \emph {et~al.}(2018)\citenamefont
  {Lucamarini}, \citenamefont {Yuan}, \citenamefont {Dynes},\ and\
  \citenamefont {Shields}}]{Lucamarini:TF2018}%
  \BibitemOpen
  \bibfield  {author} {\bibinfo {author} {\bibfnamefont {M.}~\bibnamefont
  {Lucamarini}}, \bibinfo {author} {\bibfnamefont {Z.~L.}\ \bibnamefont
  {Yuan}}, \bibinfo {author} {\bibfnamefont {J.~F.}\ \bibnamefont {Dynes}}, \
  and\ \bibinfo {author} {\bibfnamefont {A.~J.}\ \bibnamefont {Shields}},\
  }\href {https://doi.org/10.1038/s41586-018-0066-6} {\bibfield  {journal}
  {\bibinfo  {journal} {Nature}\ }\textbf {\bibinfo {volume} {557}},\ \bibinfo
  {pages} {397} (\bibinfo {year} {2018})}\BibitemShut {NoStop}%
\bibitem [{\citenamefont {Minder}\ \emph {et~al.}(2019)\citenamefont {Minder},
  \citenamefont {Pittaluga}, \citenamefont {Roberts}, \citenamefont
  {Lucamarini}, \citenamefont {Dynes}, \citenamefont {Yuan},\ and\
  \citenamefont {Shields}}]{Minder:TF2019}%
  \BibitemOpen
  \bibfield  {author} {\bibinfo {author} {\bibfnamefont {M.}~\bibnamefont
  {Minder}}, \bibinfo {author} {\bibfnamefont {M.}~\bibnamefont {Pittaluga}},
  \bibinfo {author} {\bibfnamefont {G.~L.}\ \bibnamefont {Roberts}}, \bibinfo
  {author} {\bibfnamefont {M.}~\bibnamefont {Lucamarini}}, \bibinfo {author}
  {\bibfnamefont {J.~F.}\ \bibnamefont {Dynes}}, \bibinfo {author}
  {\bibfnamefont {Z.~L.}\ \bibnamefont {Yuan}}, \ and\ \bibinfo {author}
  {\bibfnamefont {A.~J.}\ \bibnamefont {Shields}},\ }\href
  {https://doi.org/10.1038/s41566-019-0377-7} {\bibfield  {journal} {\bibinfo
  {journal} {Nature Photonics}\ }\textbf {\bibinfo {volume} {13}},\ \bibinfo
  {pages} {334–338} (\bibinfo {year} {2019})}\BibitemShut {NoStop}%
\bibitem [{\citenamefont {Pirandola}\ \emph {et~al.}(2013)\citenamefont
  {Pirandola}, \citenamefont {Ottaviani}, \citenamefont {Spedalieri},
  \citenamefont {Weedbrook},\ and\ \citenamefont
  {Braunstein}}]{Pirandola:CVMDI2013}%
  \BibitemOpen
  \bibfield  {author} {\bibinfo {author} {\bibfnamefont {S.}~\bibnamefont
  {Pirandola}}, \bibinfo {author} {\bibfnamefont {C.}~\bibnamefont
  {Ottaviani}}, \bibinfo {author} {\bibfnamefont {G.}~\bibnamefont
  {Spedalieri}}, \bibinfo {author} {\bibfnamefont {C.}~\bibnamefont
  {Weedbrook}}, \ and\ \bibinfo {author} {\bibfnamefont {S.~L.}\ \bibnamefont
  {Braunstein}},\ }\href@noop {} {\bibfield  {journal} {\bibinfo  {journal}
  {arXiv:1312.4104v1}\ } (\bibinfo {year} {2013})}\BibitemShut {NoStop}%
\bibitem [{\citenamefont {Pirandola}\ \emph {et~al.}(2015)\citenamefont
  {Pirandola}, \citenamefont {Ottaviani}, \citenamefont {Spedalieri},
  \citenamefont {Weedbrook}, \citenamefont {Braunstein}, \citenamefont {Lloyd},
  \citenamefont {Gehring}, \citenamefont {Jacobsen},\ and\ \citenamefont
  {Andersen}}]{Pirandola:CVMDI2015}%
  \BibitemOpen
  \bibfield  {author} {\bibinfo {author} {\bibfnamefont {S.}~\bibnamefont
  {Pirandola}}, \bibinfo {author} {\bibfnamefont {C.}~\bibnamefont
  {Ottaviani}}, \bibinfo {author} {\bibfnamefont {G.}~\bibnamefont
  {Spedalieri}}, \bibinfo {author} {\bibfnamefont {C.}~\bibnamefont
  {Weedbrook}}, \bibinfo {author} {\bibfnamefont {S.~L.}\ \bibnamefont
  {Braunstein}}, \bibinfo {author} {\bibfnamefont {S.}~\bibnamefont {Lloyd}},
  \bibinfo {author} {\bibfnamefont {T.}~\bibnamefont {Gehring}}, \bibinfo
  {author} {\bibfnamefont {C.~S.}\ \bibnamefont {Jacobsen}}, \ and\ \bibinfo
  {author} {\bibfnamefont {U.~L.}\ \bibnamefont {Andersen}},\ }\href {\doibase
  10.1038/nphoton.2015.83} {\bibfield  {journal} {\bibinfo  {journal} {Nature
  Photonics}\ }\textbf {\bibinfo {volume} {9}},\ \bibinfo {pages} {397}
  (\bibinfo {year} {2015})}\BibitemShut {NoStop}%
\bibitem [{\citenamefont {Ma}\ \emph {et~al.}(2014)\citenamefont {Ma},
  \citenamefont {Sun}, \citenamefont {Jiang}, \citenamefont {Gui},\ and\
  \citenamefont {Liang}}]{PhysRevA.89.042335}%
  \BibitemOpen
  \bibfield  {author} {\bibinfo {author} {\bibfnamefont {X.-C.}\ \bibnamefont
  {Ma}}, \bibinfo {author} {\bibfnamefont {S.-H.}\ \bibnamefont {Sun}},
  \bibinfo {author} {\bibfnamefont {M.-S.}\ \bibnamefont {Jiang}}, \bibinfo
  {author} {\bibfnamefont {M.}~\bibnamefont {Gui}}, \ and\ \bibinfo {author}
  {\bibfnamefont {L.-M.}\ \bibnamefont {Liang}},\ }\href {\doibase
  10.1103/PhysRevA.89.042335} {\bibfield  {journal} {\bibinfo  {journal} {Phys.
  Rev. A}\ }\textbf {\bibinfo {volume} {89}},\ \bibinfo {pages} {042335}
  (\bibinfo {year} {2014})}\BibitemShut {NoStop}%
\bibitem [{\citenamefont {Li}\ \emph {et~al.}(2014)\citenamefont {Li},
  \citenamefont {Zhang}, \citenamefont {Xu}, \citenamefont {Peng},\ and\
  \citenamefont {Guo}}]{PhysRevA.89.052301}%
  \BibitemOpen
  \bibfield  {author} {\bibinfo {author} {\bibfnamefont {Z.}~\bibnamefont
  {Li}}, \bibinfo {author} {\bibfnamefont {Y.-C.}\ \bibnamefont {Zhang}},
  \bibinfo {author} {\bibfnamefont {F.}~\bibnamefont {Xu}}, \bibinfo {author}
  {\bibfnamefont {X.}~\bibnamefont {Peng}}, \ and\ \bibinfo {author}
  {\bibfnamefont {H.}~\bibnamefont {Guo}},\ }\href {\doibase
  10.1103/PhysRevA.89.052301} {\bibfield  {journal} {\bibinfo  {journal} {Phys.
  Rev. A}\ }\textbf {\bibinfo {volume} {89}},\ \bibinfo {pages} {052301}
  (\bibinfo {year} {2014})}\BibitemShut {NoStop}%
\bibitem [{\citenamefont {Zhao}\ \emph {et~al.}(2018)\citenamefont {Zhao},
  \citenamefont {Zhang}, \citenamefont {Xu}, \citenamefont {Yu},\ and\
  \citenamefont {Guo}}]{Zhao:CVMDI-subt2018}%
  \BibitemOpen
  \bibfield  {author} {\bibinfo {author} {\bibfnamefont {Y.}~\bibnamefont
  {Zhao}}, \bibinfo {author} {\bibfnamefont {Y.}~\bibnamefont {Zhang}},
  \bibinfo {author} {\bibfnamefont {B.}~\bibnamefont {Xu}}, \bibinfo {author}
  {\bibfnamefont {S.}~\bibnamefont {Yu}}, \ and\ \bibinfo {author}
  {\bibfnamefont {H.}~\bibnamefont {Guo}},\ }\href {\doibase
  10.1103/PhysRevA.97.042328} {\bibfield  {journal} {\bibinfo  {journal} {Phys.
  Rev. A}\ }\textbf {\bibinfo {volume} {97}},\ \bibinfo {pages} {042328}
  (\bibinfo {year} {2018})}\BibitemShut {NoStop}%
\bibitem [{\citenamefont {Ma}\ \emph {et~al.}(2018)\citenamefont {Ma},
  \citenamefont {Huang}, \citenamefont {Bai}, \citenamefont {Wang},
  \citenamefont {Bao},\ and\ \citenamefont {Zeng}}]{Ma:CVMDI-subt2018}%
  \BibitemOpen
  \bibfield  {author} {\bibinfo {author} {\bibfnamefont {H.-X.}\ \bibnamefont
  {Ma}}, \bibinfo {author} {\bibfnamefont {P.}~\bibnamefont {Huang}}, \bibinfo
  {author} {\bibfnamefont {D.-Y.}\ \bibnamefont {Bai}}, \bibinfo {author}
  {\bibfnamefont {S.-Y.}\ \bibnamefont {Wang}}, \bibinfo {author}
  {\bibfnamefont {W.-S.}\ \bibnamefont {Bao}}, \ and\ \bibinfo {author}
  {\bibfnamefont {G.-H.}\ \bibnamefont {Zeng}},\ }\href {\doibase
  10.1103/PhysRevA.97.042329} {\bibfield  {journal} {\bibinfo  {journal} {Phys.
  Rev. A}\ }\textbf {\bibinfo {volume} {97}},\ \bibinfo {pages} {042329}
  (\bibinfo {year} {2018})}\BibitemShut {NoStop}%
\bibitem [{\citenamefont {Zhang}\ \emph {et~al.}(2015)\citenamefont {Zhang},
  \citenamefont {Li}, \citenamefont {Weedbrook}, \citenamefont {Marshall},
  \citenamefont {Pirandola}, \citenamefont {Yu},\ and\ \citenamefont
  {Guo}}]{Zhang:CVMDI-NLA2015}%
  \BibitemOpen
  \bibfield  {author} {\bibinfo {author} {\bibfnamefont {Y.}~\bibnamefont
  {Zhang}}, \bibinfo {author} {\bibfnamefont {Z.}~\bibnamefont {Li}}, \bibinfo
  {author} {\bibfnamefont {C.}~\bibnamefont {Weedbrook}}, \bibinfo {author}
  {\bibfnamefont {K.}~\bibnamefont {Marshall}}, \bibinfo {author}
  {\bibfnamefont {S.}~\bibnamefont {Pirandola}}, \bibinfo {author}
  {\bibfnamefont {S.}~\bibnamefont {Yu}}, \ and\ \bibinfo {author}
  {\bibfnamefont {H.}~\bibnamefont {Guo}},\ }\href {\doibase 10.3390/e17074547}
  {\bibfield  {journal} {\bibinfo  {journal} {Entropy}\ }\textbf {\bibinfo
  {volume} {17}},\ \bibinfo {pages} {4547} (\bibinfo {year}
  {2015})}\BibitemShut {NoStop}%
\bibitem [{\citenamefont {Wang}\ \emph {et~al.}(2019)\citenamefont {Wang},
  \citenamefont {Wang},\ and\ \citenamefont {Li}}]{Wang:CVMDI-NLA2019}%
  \BibitemOpen
  \bibfield  {author} {\bibinfo {author} {\bibfnamefont {P.}~\bibnamefont
  {Wang}}, \bibinfo {author} {\bibfnamefont {X.}~\bibnamefont {Wang}}, \ and\
  \bibinfo {author} {\bibfnamefont {Y.}~\bibnamefont {Li}},\ }\href
  {https://link.aps.org/doi/10.1103/PhysRevA.99.042309} {\bibfield  {journal}
  {\bibinfo  {journal} {Phys. Rev. A}\ }\textbf {\bibinfo {volume} {99}},\
  \bibinfo {pages} {042309} (\bibinfo {year} {2019})}\BibitemShut {NoStop}%
\bibitem [{\citenamefont {Ghalaii}\ and\ \citenamefont
  {Pirandola}(2023)}]{GhalaiiPRA2023}%
  \BibitemOpen
  \bibfield  {author} {\bibinfo {author} {\bibfnamefont {M.}~\bibnamefont
  {Ghalaii}}\ and\ \bibinfo {author} {\bibfnamefont {S.}~\bibnamefont
  {Pirandola}},\ }\href {\doibase 10.1103/PhysRevA.108.042621} {\bibfield
  {journal} {\bibinfo  {journal} {Phys. Rev. A}\ }\textbf {\bibinfo {volume}
  {108}},\ \bibinfo {pages} {042621} (\bibinfo {year} {2023})}\BibitemShut
  {NoStop}%
\bibitem [{\citenamefont {Zhang}\ \emph {et~al.}(2014)\citenamefont {Zhang},
  \citenamefont {Li}, \citenamefont {Yu}, \citenamefont {Gu}, \citenamefont
  {Peng},\ and\ \citenamefont {Guo}}]{Zhang:CVMDI-SqSt2014}%
  \BibitemOpen
  \bibfield  {author} {\bibinfo {author} {\bibfnamefont {Y.-C.}\ \bibnamefont
  {Zhang}}, \bibinfo {author} {\bibfnamefont {Z.}~\bibnamefont {Li}}, \bibinfo
  {author} {\bibfnamefont {S.}~\bibnamefont {Yu}}, \bibinfo {author}
  {\bibfnamefont {W.}~\bibnamefont {Gu}}, \bibinfo {author} {\bibfnamefont
  {X.}~\bibnamefont {Peng}}, \ and\ \bibinfo {author} {\bibfnamefont
  {H.}~\bibnamefont {Guo}},\ }\href {\doibase 10.1103/PhysRevA.90.052325}
  {\bibfield  {journal} {\bibinfo  {journal} {Phys. Rev. A}\ }\textbf {\bibinfo
  {volume} {90}},\ \bibinfo {pages} {052325} (\bibinfo {year}
  {2014})}\BibitemShut {NoStop}%
\bibitem [{\citenamefont {Bai}\ \emph {et~al.}(2019{\natexlab{a}})\citenamefont
  {Bai}, \citenamefont {Huang}, \citenamefont {Ma}, \citenamefont {Wang},\ and\
  \citenamefont {Zeng}}]{Bai:CVMDI-ThSt2019}%
  \BibitemOpen
  \bibfield  {author} {\bibinfo {author} {\bibfnamefont {D.}~\bibnamefont
  {Bai}}, \bibinfo {author} {\bibfnamefont {P.}~\bibnamefont {Huang}}, \bibinfo
  {author} {\bibfnamefont {H.}~\bibnamefont {Ma}}, \bibinfo {author}
  {\bibfnamefont {T.}~\bibnamefont {Wang}}, \ and\ \bibinfo {author}
  {\bibfnamefont {G.}~\bibnamefont {Zeng}},\ }\href
  {https://doi.org/10.1088/1361-6455/ab0b2a} {\bibfield  {journal} {\bibinfo
  {journal} {Journal of Physics B: Atomic, Molecular and Optical Physics}\
  }\textbf {\bibinfo {volume} {52}},\ \bibinfo {pages} {135502} (\bibinfo
  {year} {2019}{\natexlab{a}})}\BibitemShut {NoStop}%
\bibitem [{\citenamefont {Bai}\ \emph {et~al.}(2019{\natexlab{b}})\citenamefont
  {Bai}, \citenamefont {Huang}, \citenamefont {Zhu}, \citenamefont {Ma},
  \citenamefont {Xiao}, \citenamefont {Wang},\ and\ \citenamefont
  {Zeng}}]{Bai:CVMDI-UD2020}%
  \BibitemOpen
  \bibfield  {author} {\bibinfo {author} {\bibfnamefont {D.}~\bibnamefont
  {Bai}}, \bibinfo {author} {\bibfnamefont {P.}~\bibnamefont {Huang}}, \bibinfo
  {author} {\bibfnamefont {Y.}~\bibnamefont {Zhu}}, \bibinfo {author}
  {\bibfnamefont {H.}~\bibnamefont {Ma}}, \bibinfo {author} {\bibfnamefont
  {T.}~\bibnamefont {Xiao}}, \bibinfo {author} {\bibfnamefont {T.}~\bibnamefont
  {Wang}}, \ and\ \bibinfo {author} {\bibfnamefont {G.}~\bibnamefont {Zeng}},\
  }\href {https://doi.org/10.1007/s11128-019-2546-5} {\bibfield  {journal}
  {\bibinfo  {journal} {Quantum Information Processing}\ }\textbf {\bibinfo
  {volume} {19}} (\bibinfo {year} {2019}{\natexlab{b}})}\BibitemShut {NoStop}%
\bibitem [{\citenamefont {Ma}\ \emph {et~al.}(2019)\citenamefont {Ma},
  \citenamefont {Huang}, \citenamefont {Bai}, \citenamefont {Wang},
  \citenamefont {Wang}, \citenamefont {Bao},\ and\ \citenamefont
  {Zeng}}]{Ma:CVMDI-DM2019}%
  \BibitemOpen
  \bibfield  {author} {\bibinfo {author} {\bibfnamefont {H.-X.}\ \bibnamefont
  {Ma}}, \bibinfo {author} {\bibfnamefont {P.}~\bibnamefont {Huang}}, \bibinfo
  {author} {\bibfnamefont {D.-Y.}\ \bibnamefont {Bai}}, \bibinfo {author}
  {\bibfnamefont {T.}~\bibnamefont {Wang}}, \bibinfo {author} {\bibfnamefont
  {S.-Y.}\ \bibnamefont {Wang}}, \bibinfo {author} {\bibfnamefont {W.-S.}\
  \bibnamefont {Bao}}, \ and\ \bibinfo {author} {\bibfnamefont {G.-H.}\
  \bibnamefont {Zeng}},\ }\href {\doibase 10.1103/PhysRevA.99.022322}
  {\bibfield  {journal} {\bibinfo  {journal} {Phys. Rev. A}\ }\textbf {\bibinfo
  {volume} {99}},\ \bibinfo {pages} {022322} (\bibinfo {year}
  {2019})}\BibitemShut {NoStop}%
\bibitem [{\citenamefont {Liao}\ \emph {et~al.}(2018)\citenamefont {Liao},
  \citenamefont {Wang}, \citenamefont {Huang},\ and\ \citenamefont
  {Guo}}]{Liao:CVMDI-DPhM2018}%
  \BibitemOpen
  \bibfield  {author} {\bibinfo {author} {\bibfnamefont {Q.}~\bibnamefont
  {Liao}}, \bibinfo {author} {\bibfnamefont {Y.}~\bibnamefont {Wang}}, \bibinfo
  {author} {\bibfnamefont {D.}~\bibnamefont {Huang}}, \ and\ \bibinfo {author}
  {\bibfnamefont {Y.}~\bibnamefont {Guo}},\ }\href {\doibase
  10.1364/OE.26.019907} {\bibfield  {journal} {\bibinfo  {journal} {Opt.
  Express}\ }\textbf {\bibinfo {volume} {26}},\ \bibinfo {pages} {19907}
  (\bibinfo {year} {2018})}\BibitemShut {NoStop}%
\bibitem [{\citenamefont {Papanastasiou}\ \emph {et~al.}(2017)\citenamefont
  {Papanastasiou}, \citenamefont {Ottaviani},\ and\ \citenamefont
  {Pirandola}}]{Papanastasiou:FSCVMDI2017}%
  \BibitemOpen
  \bibfield  {author} {\bibinfo {author} {\bibfnamefont {P.}~\bibnamefont
  {Papanastasiou}}, \bibinfo {author} {\bibfnamefont {C.}~\bibnamefont
  {Ottaviani}}, \ and\ \bibinfo {author} {\bibfnamefont {S.}~\bibnamefont
  {Pirandola}},\ }\href {\doibase 10.1103/PhysRevA.96.042332} {\bibfield
  {journal} {\bibinfo  {journal} {Phys. Rev. A}\ }\textbf {\bibinfo {volume}
  {96}},\ \bibinfo {pages} {042332} (\bibinfo {year} {2017})}\BibitemShut
  {NoStop}%
\bibitem [{\citenamefont {Mountogiannakis}\ \emph {et~al.}(2022)\citenamefont
  {Mountogiannakis}, \citenamefont {Papanastasiou},\ and\ \citenamefont
  {Pirandola}}]{Mountogiannakis:CompFS2022}%
  \BibitemOpen
  \bibfield  {author} {\bibinfo {author} {\bibfnamefont {A.~G.}\ \bibnamefont
  {Mountogiannakis}}, \bibinfo {author} {\bibfnamefont {P.}~\bibnamefont
  {Papanastasiou}}, \ and\ \bibinfo {author} {\bibfnamefont {S.}~\bibnamefont
  {Pirandola}},\ }\href {\doibase 10.1103/PhysRevA.106.042606} {\bibfield
  {journal} {\bibinfo  {journal} {Phys. Rev. A}\ }\textbf {\bibinfo {volume}
  {106}},\ \bibinfo {pages} {042606} (\bibinfo {year} {2022})}\BibitemShut
  {NoStop}%
\bibitem [{\citenamefont {Wilkinson}\ \emph {et~al.}(2020)\citenamefont
  {Wilkinson}, \citenamefont {Papanastasiou}, \citenamefont {Ottaviani},
  \citenamefont {Gehring},\ and\ \citenamefont
  {Pirandola}}]{Wilkinson:CVMDI2020}%
  \BibitemOpen
  \bibfield  {author} {\bibinfo {author} {\bibfnamefont {K.~N.}\ \bibnamefont
  {Wilkinson}}, \bibinfo {author} {\bibfnamefont {P.}~\bibnamefont
  {Papanastasiou}}, \bibinfo {author} {\bibfnamefont {C.}~\bibnamefont
  {Ottaviani}}, \bibinfo {author} {\bibfnamefont {T.}~\bibnamefont {Gehring}},
  \ and\ \bibinfo {author} {\bibfnamefont {S.}~\bibnamefont {Pirandola}},\
  }\href {\doibase 10.1103/PhysRevResearch.2.033424} {\bibfield  {journal}
  {\bibinfo  {journal} {Phys. Rev. Research}\ }\textbf {\bibinfo {volume}
  {2}},\ \bibinfo {pages} {033424} (\bibinfo {year} {2020})}\BibitemShut
  {NoStop}%
\bibitem [{\citenamefont {Fletcher}\ and\ \citenamefont
  {Pirandola}(2022)}]{Fletcher:3nodeCVMDI2022}%
  \BibitemOpen
  \bibfield  {author} {\bibinfo {author} {\bibfnamefont {A.~I.}\ \bibnamefont
  {Fletcher}}\ and\ \bibinfo {author} {\bibfnamefont {S.}~\bibnamefont
  {Pirandola}},\ }\href@noop {} {\bibfield  {journal} {\bibinfo  {journal} {Sci
  Rep}\ }\textbf {\bibinfo {volume} {12}},\ \bibinfo {pages} {17329} (\bibinfo
  {year} {2022})}\BibitemShut {NoStop}%
\bibitem [{\citenamefont {Ghalaii}\ \emph {et~al.}(2022)\citenamefont
  {Ghalaii}, \citenamefont {Papanastasiou},\ and\ \citenamefont
  {Pirandola}}]{Ghalaii:GQNets2022}%
  \BibitemOpen
  \bibfield  {author} {\bibinfo {author} {\bibfnamefont {M.}~\bibnamefont
  {Ghalaii}}, \bibinfo {author} {\bibfnamefont {P.}~\bibnamefont
  {Papanastasiou}}, \ and\ \bibinfo {author} {\bibfnamefont {S.}~\bibnamefont
  {Pirandola}},\ }\href@noop {} {\bibfield  {journal} {\bibinfo  {journal} {npj
  Quantum Inf}\ }\textbf {\bibinfo {volume} {8}},\ \bibinfo {pages} {105}
  (\bibinfo {year} {2022})}\BibitemShut {NoStop}%
\bibitem [{\citenamefont {Garc\'{\i}a-Patr\'on}\ and\ \citenamefont
  {Cerf}(2006)}]{GarciaPatron:Optim2006}%
  \BibitemOpen
  \bibfield  {author} {\bibinfo {author} {\bibfnamefont {R.}~\bibnamefont
  {Garc\'{\i}a-Patr\'on}}\ and\ \bibinfo {author} {\bibfnamefont {N.~J.}\
  \bibnamefont {Cerf}},\ }\href {\doibase 10.1103/PhysRevLett.97.190503}
  {\bibfield  {journal} {\bibinfo  {journal} {Phys. Rev. Lett.}\ }\textbf
  {\bibinfo {volume} {97}},\ \bibinfo {pages} {190503} (\bibinfo {year}
  {2006})}\BibitemShut {NoStop}%
\bibitem [{\citenamefont {Fletcher}(2023)}]{Fletcher:thesis2023}%
  \BibitemOpen
  \bibfield  {author} {\bibinfo {author} {\bibfnamefont {A.}~\bibnamefont
  {Fletcher}},\ }\href@noop {} {\bibfield  {journal} {\bibinfo  {journal} {PhD
  Thesis, University of York}\ } (\bibinfo {year} {2023})}\BibitemShut
  {NoStop}%
\bibitem [{\citenamefont {Silberhorn}\ \emph {et~al.}(2002)\citenamefont
  {Silberhorn}, \citenamefont {Ralph}, \citenamefont {L{\"{u}}tkenhaus},\ and\
  \citenamefont {Leuchs}}]{Silberhorn:3db2002}%
  \BibitemOpen
  \bibfield  {author} {\bibinfo {author} {\bibfnamefont {C.}~\bibnamefont
  {Silberhorn}}, \bibinfo {author} {\bibfnamefont {T.~C.}\ \bibnamefont
  {Ralph}}, \bibinfo {author} {\bibfnamefont {N.}~\bibnamefont
  {L{\"{u}}tkenhaus}}, \ and\ \bibinfo {author} {\bibfnamefont
  {G.}~\bibnamefont {Leuchs}},\ }\href {\doibase 10.1103/PhysRevLett.89.167901}
  {\bibfield  {journal} {\bibinfo  {journal} {Physical Review Letters}\
  }\textbf {\bibinfo {volume} {89}},\ \bibinfo {pages} {167901} (\bibinfo
  {year} {2002})}\BibitemShut {NoStop}%
\bibitem [{\citenamefont {Lupo}\ \emph {et~al.}(2018)\citenamefont {Lupo},
  \citenamefont {Ottaviani}, \citenamefont {Papanastasiou},\ and\ \citenamefont
  {Pirandola}}]{Lupo:FSCVMDI2018}%
  \BibitemOpen
  \bibfield  {author} {\bibinfo {author} {\bibfnamefont {C.}~\bibnamefont
  {Lupo}}, \bibinfo {author} {\bibfnamefont {C.}~\bibnamefont {Ottaviani}},
  \bibinfo {author} {\bibfnamefont {P.}~\bibnamefont {Papanastasiou}}, \ and\
  \bibinfo {author} {\bibfnamefont {S.}~\bibnamefont {Pirandola}},\ }\href
  {\doibase 10.1103/PhysRevA.97.052327} {\bibfield  {journal} {\bibinfo
  {journal} {Phys. Rev. A}\ }\textbf {\bibinfo {volume} {97}},\ \bibinfo
  {pages} {052327} (\bibinfo {year} {2018})}\BibitemShut {NoStop}%
\bibitem [{\citenamefont {Papanastasiou}\ \emph {et~al.}(2023)\citenamefont
  {Papanastasiou}, \citenamefont {Mountogiannakis},\ and\ \citenamefont
  {Pirandola}}]{Papanastasiou:CVMDI23}%
  \BibitemOpen
  \bibfield  {author} {\bibinfo {author} {\bibfnamefont {P.}~\bibnamefont
  {Papanastasiou}}, \bibinfo {author} {\bibfnamefont {A.}~\bibnamefont
  {Mountogiannakis}}, \ and\ \bibinfo {author} {\bibfnamefont {S.}~\bibnamefont
  {Pirandola}},\ }\href {\doibase 10.1038/s41598-023-37699-5} {\bibfield
  {journal} {\bibinfo  {journal} {Sci. Rep.}\ }\textbf {\bibinfo {volume}
  {13}},\ \bibinfo {pages} {11636} (\bibinfo {year} {2023})}\BibitemShut
  {NoStop}%
\bibitem [{\citenamefont {Pirandola}\ and\ \citenamefont
  {Papanastasiou}(2024)}]{Pirandola:QKDrates2024}%
  \BibitemOpen
  \bibfield  {author} {\bibinfo {author} {\bibfnamefont {S.}~\bibnamefont
  {Pirandola}}\ and\ \bibinfo {author} {\bibfnamefont {P.}~\bibnamefont
  {Papanastasiou}},\ }\href {\doibase 10.1103/PhysRevResearch.6.023321}
  {\bibfield  {journal} {\bibinfo  {journal} {Phys. Rev. Res.}\ }\textbf
  {\bibinfo {volume} {6}},\ \bibinfo {pages} {023321} (\bibinfo {year}
  {2024})}\BibitemShut {NoStop}%
\bibitem [{\citenamefont {Tian}\ \emph {et~al.}(2022)\citenamefont {Tian},
  \citenamefont {Wang}, \citenamefont {Liu}, \citenamefont {Du}, \citenamefont
  {Liu}, \citenamefont {Lu}, \citenamefont {Wang},\ and\ \citenamefont
  {Li}}]{tian2022experimental}%
  \BibitemOpen
  \bibfield  {author} {\bibinfo {author} {\bibfnamefont {Y.}~\bibnamefont
  {Tian}}, \bibinfo {author} {\bibfnamefont {P.}~\bibnamefont {Wang}}, \bibinfo
  {author} {\bibfnamefont {J.}~\bibnamefont {Liu}}, \bibinfo {author}
  {\bibfnamefont {S.}~\bibnamefont {Du}}, \bibinfo {author} {\bibfnamefont
  {W.}~\bibnamefont {Liu}}, \bibinfo {author} {\bibfnamefont {Z.}~\bibnamefont
  {Lu}}, \bibinfo {author} {\bibfnamefont {X.}~\bibnamefont {Wang}}, \ and\
  \bibinfo {author} {\bibfnamefont {Y.}~\bibnamefont {Li}},\ }\href@noop {}
  {\bibfield  {journal} {\bibinfo  {journal} {Optica}\ }\textbf {\bibinfo
  {volume} {9}},\ \bibinfo {pages} {492} (\bibinfo {year} {2022})}\BibitemShut
  {NoStop}%
\bibitem [{\citenamefont {Hajomer}\ \emph {et~al.}(2022)\citenamefont
  {Hajomer}, \citenamefont {Nguyen},\ and\ \citenamefont
  {Gehring}}]{hajomer2022high}%
  \BibitemOpen
  \bibfield  {author} {\bibinfo {author} {\bibfnamefont {A.~A.}\ \bibnamefont
  {Hajomer}}, \bibinfo {author} {\bibfnamefont {H.~Q.}\ \bibnamefont {Nguyen}},
  \ and\ \bibinfo {author} {\bibfnamefont {T.}~\bibnamefont {Gehring}},\
  }\href@noop {} {\bibfield  {journal} {\bibinfo  {journal} {arXiv preprint
  arXiv:2210.07576}\ } (\bibinfo {year} {2022})}\BibitemShut {NoStop}%
\bibitem [{\citenamefont {Hajomer}\ \emph {et~al.}(2025)\citenamefont
  {Hajomer}, \citenamefont {Andersen},\ and\ \citenamefont
  {Gehring}}]{hajomer2023real}%
  \BibitemOpen
  \bibfield  {author} {\bibinfo {author} {\bibfnamefont {A.~A.}\ \bibnamefont
  {Hajomer}}, \bibinfo {author} {\bibfnamefont {U.~L.}\ \bibnamefont
  {Andersen}}, \ and\ \bibinfo {author} {\bibfnamefont {T.}~\bibnamefont
  {Gehring}},\ }\href@noop {} {\bibfield  {journal} {\bibinfo  {journal} {arXiv
  preprint arXiv:2303.01611}\ } (\bibinfo {year} {2025})}\BibitemShut {NoStop}%
\bibitem [{\citenamefont {Hajomer}\ \emph {et~al.}(2024)\citenamefont
  {Hajomer}, \citenamefont {Bruynsteen}, \citenamefont {Derkach}, \citenamefont
  {Jain}, \citenamefont {Bomhals}, \citenamefont {Bastiaens}, \citenamefont
  {Andersen}, \citenamefont {Yin},\ and\ \citenamefont
  {Gehring}}]{hajomer2024continuous}%
  \BibitemOpen
  \bibfield  {author} {\bibinfo {author} {\bibfnamefont {A.~A.}\ \bibnamefont
  {Hajomer}}, \bibinfo {author} {\bibfnamefont {C.}~\bibnamefont {Bruynsteen}},
  \bibinfo {author} {\bibfnamefont {I.}~\bibnamefont {Derkach}}, \bibinfo
  {author} {\bibfnamefont {N.}~\bibnamefont {Jain}}, \bibinfo {author}
  {\bibfnamefont {A.}~\bibnamefont {Bomhals}}, \bibinfo {author} {\bibfnamefont
  {S.}~\bibnamefont {Bastiaens}}, \bibinfo {author} {\bibfnamefont {U.~L.}\
  \bibnamefont {Andersen}}, \bibinfo {author} {\bibfnamefont {X.}~\bibnamefont
  {Yin}}, \ and\ \bibinfo {author} {\bibfnamefont {T.}~\bibnamefont
  {Gehring}},\ }\href@noop {} {\bibfield  {journal} {\bibinfo  {journal}
  {Optica}\ }\textbf {\bibinfo {volume} {11}},\ \bibinfo {pages} {1197}
  (\bibinfo {year} {2024})}\BibitemShut {NoStop}%
\end{thebibliography}%

\end{document}